# Critically Evaluated Energy Levels, Wavelengths, Transition Probabilities, and Intensities of Six-Times Ionized Cesium: Cs VII

Abid Husain, K. Haris*, S. Jabeen, and A. Tauheed
Department of Physics, Aligarh Muslim University, Aligarh, UP-202002, India
*Corresponding Author: kharisphy@gmail.com

**Abstract**
Previously reported works on the spectrum of Cs VII are critically studied using supplementary spectrograms recorded on a 3-m normal incidence vacuum spectrograph in the wavelength region 300–1240 Å at the Antigonish laboratory (Canada). We confirmed the results of the earlier work of Gayasov and Joshi on this spectrum. Our analysis is supported by extended calculations with the pseudo-relativistic Hartree-Fock (HFR) method with superposition of configuration interactions implemented in Cowan's suite of codes. In this critical evaluation, in addition to the accurate energy levels of Cs VII with their uncertainties, observed and Ritz wavelengths with uncertainties and transition probabilities, the uniformly-scaled intensities of Cs VII lines are also presented. A total of 196 lines attributed to 197 transitions enabled us to optimize the energy values of 72 levels in Cs VII spectrum. Furthermore, Ritz wavelengths of 141 possibly observable lines are provided along with their transition probabilities.

## 1. Introduction

With the advent of lasers and atom/ion traps, our scientific interest in highly charged ions (HCIs) has been growing, as they present an enhanced avenue for studying the electron-electron correlations, quantum electrodynamics (QED), hyperfine, relativistic and finite-nuclear-size effects [1]. Additionally, several HCIs in the indium (In I) sequence were proposed as suitable candidates for the future development of new generation optical clocks [2–4]. These ultra-precise clocks use narrow lines of either parity- or *J*- forbidden transitions. They are expected to have a fractional frequency uncertainty on the level of $10^{-19}$ or smaller. Therefore, these clocks have potential applications in testing new physics, such as the study of variation of fundamental constants, in development of highly sensitive quantum-based tools for geodesy, very-long-baseline interferometry for telescope array synchronization, climate change studies, hydrology, deep-space applications, and inertial navigation [5,6]. Recently, Safronova et al. [3,4] had calculated the energies of the $5s^25p$–$5s^24f$ clock transitions and their $\alpha$-variation sensitivity coefficients for many In-like clock ions ($Ce^{9+}$, $Pr^{10+}$, $Nd^{11+}$). However, due to scarcity of experimental data on these ions, their theoretical results were compared with experimental data present for the lower members ($Cs^{6+}$, $Ba^{7+}$) of the sequence. This is one of the (indirect) applications of experimental atomic data, which requires some quality checks to be made before their usage.

The ground configuration of Cs VII is $5s^25p$ with $^2P^o_{1/2,\,3/2}$ levels, and the excited configurations are of the type $5s^2n\ell$ ($n$≥4; $\ell$=s, p, d, f, ….), $5s5p^2$, $5p^3$, $5s5pn\ell$ ($n$≥4; $\ell$=s, p, d, f, ….), and $5p^25d$. The first analysis of this spectrum was carried out by Kaufman and Sugar [7]; they reported the energy levels of the $5s^25p$, $5s5p^2$, $5s^25d$, and $5s^26s$ configurations. Altogether, ten energy levels were established with the help of fourteen observed lines in the 400–700 Å wavelength region. Only doublet ($^2D$, $^2P$, and $^2S$) levels belonging to the $5s5p^2$ configuration were given in Ref. [7]; the $^4P$ levels of this configuration were left out. The energy values of these $^4P$ levels were later reported by Tauheed et al. in 1992 [8]. In 1999, Gayasov and Joshi [9] not only confirmed all the previously reported levels but also extended the work to incorporate new configurations $5s^26p$, $5s^2(4f+5f)$, $5p^3$, 5s5p(4f+5d+6s). They classified 184 lines establishing 71 levels, and also reported the ionization potential of Cs VII at 687300(500) $cm^{-1}$ or 85.20(6) eV. All these updates were included in the recent compilation of cesium and its ions (Cs I–LV) by Sansonetti [10]; subsequently, those were inserted into Atomic Spectra Database (ASD) of the National Institute of Standards and Technology (NIST) [11].

In this work, one of our aims is to re-investigate the analysis of Gayasov and Joshi [9] on Cs VII using additional spectrograms of cesium recorded on a 3 m normal incidence vacuum spectrograph in the wavelength region 359–1240 Å at Antigonish laboratory (Canada). We also carried out a critical evaluation of all data on the Cs VII spectrum. As a result of this evaluation, we provide the wavelengths of observed lines (together with their Ritz counterpart), as well as the optimized energy levels with their uncertainties, uniformly-scaled intensities, and transition probabilities for observed and possibly observable lines.

## 2. Experimental details

The spectrum of cesium used for this work was recorded in the 300–1240 Å wavelength region on a 3 m normal incidence vacuum spectrograph at the Antigonish laboratory (Canada). The spectrograph was equipped with a 2400 lines per mm holographic grating of having a reciprocal dispersion of about 1.38 Å $mm^{-1}$ in the first order of diffraction. Aluminum electrodes were used in a triggered spark source. These electrodes had cavities in which a pure $CsNO_3$ or $CsCO_3$ powder was packed. The charging potential across a low-inductance capacitor (14.3 µF, 25 nH) was varied from 4 to 12 kV in discharges triggered by a 30 kV pulse. The exposures were taken on Kodak short-wave-radiation (SWR) plates. Several tracks of exposures (as many as five) were taken at varying experimental conditions: either reducing discharge voltage or inserting inductor coils in series with the spark circuit. The higher ionizations appear close to the anode, producing short lines on the plate. An inductance coil inserted in series in the discharge circuit helped to quench the higher stages of ionization on some tracks [12]. This procedure was helpful to discern the ionization states responsible for the lines. Sometimes, spatial variation of intensity of the line image was also taken into account, i.e. line characters such as polarity of the lines (short length and proximity to the anode or cathode). The spectral lines of Cs IV–Cs X have been identified by observing their characteristic appearance on various tracks of the plate. The relative positions of spectral lines along with their relative intensities were measured on a Zeiss Abbe comparator at Aligarh, and the spectrograms were calibrated with a polynomials of order 2 or 3 using lines of common impurities: carbon (C II – IV), nitrogen (N II), oxygen (O II–V), aluminum (Al II–III), and silicon (Si II–IV), together with additional lines of cesium (Cs IV–X). The wavelengths of these well-known lines were obtained from the NIST ASD [11]. The uncertainty of our wavelength measurement is estimated to be 0.010 Å for the entire region. Unlike the precisely measured wavelengths, the line intensities are visual estimates representing the blackening of the photographic emulsion. We further modelled these intensities to bring them on a uniform scale (see Section 3.4). It is important to mention here that the plates used in this work were recorded at different experimental conditions on the same instrument: the first plate covers the wavelength region 300–890 Å, while the second one is in the 730–1240 Å range of wavelength. In either case, the track with the strongest Cs VII lines was used to measure the wavelengths and the remaining tracks on each plate were used to accomplish the ionization separation.

## 3. Results and discussion

The main results of our work on Cs VII are summarized in two tables; **Table 1** contains the classified lines, and **Table 2** describes the optimized energy levels with their *LS* compositions. The details of the analysis are discussed in the sections below.

Table 1.Classified lines in Cs VII

| $I_{Obs}$[a] (arb. u.) | $\lambda_{Obs}$[b], (Å) | $\sigma_{Obs}$, $cm^{-1}$ | $\lambda_{Ritz}$[b], (Å) | $\delta\lambda_{O-c\ Ritz}$ (mÅ) | Classification | | | | $E_{low}$, $cm^{-1}$ | $E_{upp}$, $cm^{-1}$ | $gA$[d], $s^{-1}$ | CF[d] | Line Ref.[e] |
|---|---|---|---|---|---|---|---|---|---|---|---|---|---|
| | | | | | Lower level | | Upper level | | | | | | |
| | | | 308.7051(19) | | $5s^25p$ | $^2P°_{1/2}$ | $5s5p(^3P°)4f$ | $^2D_{3/2}$ | 0.0 | 323933.7 | 1.67e+09 | 0.11 | |
| | | | 319.307(3) | | $5s5p^2(^3P)$ | $^4P_{5/2}$ | $5s5p(^1P°)6s$ | $^2P°_{3/2}$ | 122261.3 | 435440 | 3.87e+08 | 0.02 | |
| 710 | 327.034(5) | 305779 | 327.0345(18) | -1 | $5s^25p$ | $^2P°_{3/2}$ | $5s5p(^3P°)4f$ | $^2D_{5/2}$ | 19379.30 | 325157.4 | 3.67e+09 | 0.13 | GJ |
| | | | 328.3486(21) | | $5s^25p$ | $^2P°_{3/2}$ | $5s5p(^3P°)4f$ | $^2D_{3/2}$ | 19379.30 | 323933.7 | 3.08e+08 | 0.11 | |
| | | | 330.908(4) | | $5s^25p$ | $^2P°_{1/2}$ | $5s5p(^1P°)4f$ | $^2D_{3/2}$ | 0.0 | 302199 | 2.59e+09 | 0.17 | |
| | | | 330.9161(18) | | $5s5p^2(^3P)$ | $^4P_{1/2}$ | $5s5p(^3P°)6s$ | $^2P°_{3/2}$ | 104226.2 | 406417.6 | 1.65e+08 | 0.01 | |
| | | | 339.823(4) | | $5s5p^2(^1D)$ | $^2D_{3/2}$ | $5s5p(^1P°)6s$ | $^2P°_{3/2}$ | 141168.5 | 435440 | 7.99e+09 | 0.28 | |
| | | | 340.3802(19) | | $5s^25p$ | $^2P°_{3/2}$ | $5s5p(^3P°)4f$ | $^2F_{5/2}$ | 19379.30 | 313168.4 | 2.78e+08 | 0.14 | |
| | | | 341.4192(22) | | $5s5p^2(^1D)$ | $^2D_{3/2}$ | $5s5p(^1P°)6s$ | $^2P°_{1/2}$ | 141168.5 | 434063.6 | 1.49e+10 | 0.31 | |
| | | | 342.2750(19) | | $5s5p^2(^3P)$ | $^4P_{3/2}$ | $5s5p(^3P°)6s$ | $^2P°_{3/2}$ | 114254.9 | 406417.6 | 2.21e+09 | 0.58 | |
| | | | 346.912(4) | | $5s5p^2(^1D)$ | $^2D_{5/2}$ | $5s5p(^1P°)6s$ | $^2P°_{3/2}$ | 147182.2 | 435440 | 1.19e+10 | 0.17 | |
| | | | 348.709(3) | | $5s5p^2(^3P)$ | $^4P_{1/2}$ | $5s5p(^3P°)6s$ | $^2P°_{1/2}$ | 104226.2 | 390998.4 | 4.87e+08 | 0.10 | |
| 440 | 351.660(5) | 284366 | 351.660(3) | 0 | $5s5p^2(^3P)$ | $^4P_{3/2}$ | $5s5p(^3P°)6s$ | $^4P°_{5/2}$ | 114254.9 | 398620.2 | 3.19e+10 | 0.76 | GJ |
| | | | 351.9190(21) | | $5s5p^2(^3P)$ | $^4P_{5/2}$ | $5s5p(^3P°)6s$ | $^2P°_{3/2}$ | 122261.3 | 406417.6 | 4.60e+09 | 0.13 | |
| 590 | 353.582(5) | 282820 | 353.582(5) | | $5s^25p$ | $^2P°_{3/2}$ | $5s5p(^1P°)4f$ | $^2D_{3/2}$ | 19379.30 | 302199 | 5.94e+08 | 0.20 | GJ |
| 460 | 357.588(5) | 279651 | 357.589(3) | -1 | $5s5p^2(^3P)$ | $^4P_{1/2}$ | $5s5p(^3P°)6s$ | $^4P°_{3/2}$ | 104226.2 | 383877.1 | 2.83e+10 | 0.67 | GJ |
| | | | 360.557(4) | | $5s5p^2(^3P)$ | $^2P_{1/2}$ | $5s5p(^1P°)6s$ | $^2P°_{3/2}$ | 158091.3 | 435440 | 1.26e+10 | 0.42 | |
| | | | 361.345(3) | | $5s5p^2(^3P)$ | $^4P_{3/2}$ | $5s5p(^3P°)6s$ | $^2P°_{1/2}$ | 114254.9 | 390998.4 | 2.46e+08 | 0.05 | |
| 300 | 361.679(5) | 276488 | 361.6775(20) | 1 | $5s^25p$ | $^2P°_{3/2}$ | $5s5p(^1P°)4f$ | $^2D_{5/2}$ | 19379.30 | 295868.7 | 2.23e+09 | 0.17 | GJ |
| 580 | 361.848(5) | 276359 | 361.848(3) | 0 | $5s5p^2(^3P)$ | $^4P_{5/2}$ | $5s5p(^3P°)6s$ | $^4P°_{5/2}$ | 122261.3 | 398620.2 | 5.78e+10 | 0.76 | GJ |
| | | | 362.3552(24) | | $5s5p^2(^3P)$ | $^2P_{1/2}$ | $5s5p(^1P°)6s$ | $^2P°_{1/2}$ | 158091.3 | 434063.6 | 2.78e+09 | 0.10 | |
| 160 | 363.812(5) | 274867 | 363.810(3) | 2 | $5s5p^2(^3P)$ | $^4P_{1/2}$ | $5s5p(^3P°)6s$ | $^4P°_{1/2}$ | 104226.2 | 379095 | 5.83e+09 | 0.64 | GJ |
| 2300 | 365.834(5) | 273348 | 365.829(2) | 5 | $5s^25p$ | $^2P°_{1/2}$ | $5s^26s$ | $^2S_{1/2}$ | 0.0 | 273351.5 | 1.80e+10 | 0.91 | GJ |
| 180 | 370.890(5) | 269622 | 370.889(3) | 1 | $5s5p^2(^3P)$ | $^4P_{3/2}$ | $5s5p(^3P°)6s$ | $^4P°_{3/2}$ | 114254.9 | 383877.1 | 6.41e+09 | 0.52 | GJ |
| 200 | 372.902(5) | 268167 | 372.9025(22) | -1 | $5s^25p$ | $^2P°_{1/2}$ | $5s5p(^3P°)4f$ | 4F$_{3/2}$ | 0.0 | 268166.6 | 5.01e+08 | 0.30 | GJ |
| 210 | 373.923(5) | 267435 | 373.9310(22) | -8 | $5s^25p$ | $^2P°_{3/2}$ | $5s5p(^3P°)4f$ | $^4D_{5/2}$ | 19379.30 | 286808.3 | 3.39e+08 | 0.21 | GJ |
| | | | 377.0041(23) | | $5s5p^2(^1D)$ | $^2D_{3/2}$ | $5s5p(^3P°)6s$ | $^2P°_{3/2}$ | 141168.5 | 406417.6 | 3.44e+09 | 0.16 | |
| 880 | 377.584(5) | 264842 | 377.586(4) | -2 | $5s5p^2(^3P)$ | $^4P_{3/2}$ | $5s5p(^3P°)6s$ | $^4P°_{1/2}$ | 114254.9 | 379095 | 2.33e+10 | 0.69 | GJ |
| | | | 378.7321(12) | | $5s5p^2(^3P)$ | $^4P_{1/2}$ | $5s5p(^1P°)5d$ | $^2P°_{3/2}$ | 104226.2 | 368265.1 | 2.09e+08 | 0.01 | |
| | | | 380.5313(24) | | $5s5p^2(^3P)$ | $^4P_{5/2}$ | $5s^25f$ | $^2F°_{7/2}$ | 122261.3 | 385051.8 | 8.03e+07 | 0.00 | |
| 920 | 382.240(5) | 261616 | 382.240(3) | 0 | $5s5p^2(^3P)$ | $^4P_{5/2}$ | $5s5p(^3P°)6s$ | $^4P°_{3/2}$ | 122261.3 | 383877.1 | 2.96e+10 | 0.68 | GJ |
| | | | 383.4891(15) | | $5s5p^2(^3P)$ | $^4P_{1/2}$ | $5s5p(^1P°)5d$ | $^2P°_{1/2}$ | 104226.2 | 364989.8 | 1.48e+08 | 0.01 | |
| 890 | 385.751(5) | 259235 | 385.7498(24) | 1 | $5s5p^2(^1D)$ | $^2D_{5/2}$ | $5s5p(^3P°)6s$ | $^2P°_{3/2}$ | 147182.2 | 406417.6 | 6.06e+10 | 0.71 | GJ |
| | | | 388.422(3) | | $5s5p^2(^1D)$ | $^2D_{3/2}$ | $5s5p(^3P°)6s$ | $^4P°_{5/2}$ | 141168.5 | 398620.2 | 2.02e+08 | 0.66 | |
| | | | 390.472(5) | | $5s5p^2(^3P)$ | $^2P_{3/2}$ | $5s5p(^1P°)6s$ | $^2P°_{3/2}$ | 179339.4 | 435440 | 3.64e+10 | 0.63 | |
| | | | 390.819(3) | | $5s5p^2(^1S)$ | $^2S_{1/2}$ | $5s5p(^1P°)6s$ | $^2P°_{1/2}$ | 178190.9 | 434063.6 | 1.23e+10 | 0.67 | |

| $I_{Obs}$[a] (arb. u.) | $\lambda_{Obs}$[b], (Å) | $\sigma_{Obs}$, cm$^{-1}$ | $\lambda_{Ritz}$[b], (Å) | $\delta\lambda_{O-Ritz}$ (mÅ) | Classification | | | | $E_{low}$, cm$^{-1}$ | $E_{upp}$, cm$^{-1}$ | $gA$[d], s$^{-1}$ | CF[d] | Line Ref.[e] |
|---|---|---|---|---|---|---|---|---|---|---|---|---|---|
| | | | | | Lower level | | Upper level | | | | | | |
| 140 | 391.278(5) | 255573 | 391.2801(23) | -2 | $5s^{2}5p$ | $^{2}P^{\circ}_{3/2}$ | $5s5p(^{1}P^{\circ})4f$ | $^{2}F_{5/2}$ | 19379.30 | 274950.7 | 4.01e+08 | 0.20 | GJ |
| 90p | 392.566(10) | 254734 | 392.581(3) | -16 | $5s5p^{2}(^{3}P)$ | $^{2}P_{3/2}$ | $5s5p(^{1}P^{\circ})6s$ | $^{2}P^{\circ}_{1/2}$ | 179339.4 | 434063.6 | 8.91e+09 | 0.54 | TW |
| 6100 | 393.745(5) | 253971 | 393.744(3) | 1 | $5s^{2}5p$ | $^{2}P^{\circ}_{3/2}$ | $5s^{2}6s$ | $^{2}S_{1/2}$ | 19379.30 | 273351.5 | 2.88e+10 | 0.91 | GJ |
| | | | 396.2668(20) | | $5s5p^{2}(^{3}P)$ | $^{4}P_{3/2}$ | $5s5p(^{1}P^{\circ})5d$ | $^{2}D^{\circ}_{5/2}$ | 114254.9 | 366610.1 | 2.95e+08 | 0.02 | |
| | | | 397.1508(15) | | $5s5p^{2}(^{3}P)$ | $^{4}P_{1/2}$ | $5s5p(^{3}P^{\circ})5d$ | $^{2}P^{\circ}_{1/2}$ | 104226.2 | 356019.7 | 4.68e+07 | 0.00 | |
| 170 | 397.713(5) | 251438 | 397.712(3) | 1 | $5s5p^{2}(^{1}D)$ | $^{2}D_{5/2}$ | $5s5p(^{3}P^{\circ})6s$ | $^{4}P^{\circ}_{5/2}$ | 147182.2 | 398620.2 | 7.84e+09 | 0.72 | GJ |
| | | | 398.7716(12) | | $5s5p^{2}(^{3}P)$ | $^{4}P_{3/2}$ | $5s5p(^{1}P^{\circ})5d$ | $^{2}D^{\circ}_{3/2}$ | 114254.9 | 365025.0 | 2.60e+08 | 0.02 | |
| 540 | 400.272(5) | 249830 | 400.272(3) | 0 | $5s5p^{2}(^{1}D)$ | $^{2}D_{3/2}$ | $5s5p(^{3}P^{\circ})6s$ | $^{2}P^{\circ}_{1/2}$ | 141168.5 | 390998.4 | 2.77e+10 | 0.50 | GJ |
| 250 | 401.179(5) | 249265 | 401.1703(24) | 9 | $5s^{2}5p$ | $^{2}P^{\circ}_{3/2}$ | $5s5p(^{3}P^{\circ})4f$ | $^{4}F_{5/2}$ | 19379.30 | 268650.0 | 3.03e+08 | 0.30 | GJ |
| | | | 401.950(2) | | $5s^{2}5p$ | $^{2}P^{\circ}_{3/2}$ | $5s5p(^{3}P^{\circ})4f$ | $^{4}F_{3/2}$ | 19379.30 | 268166.6 | 1.08e+08 | 0.33 | |
| | | | 402.696(3) | | $5s5p^{2}(^{3}P)$ | $^{2}P_{1/2}$ | $5s5p(^{3}P^{\circ})6s$ | $^{2}P^{\circ}_{3/2}$ | 158091.3 | 406417.6 | 2.89e+09 | 0.12 | |
| | | | 406.4978(13) | | $5s5p^{2}(^{3}P)$ | $^{4}P_{5/2}$ | $5s5p(^{1}P^{\circ})5d$ | $^{2}P^{\circ}_{3/2}$ | 122261.3 | 368265.1 | 4.89e+08 | 0.02 | |
| | | | 409.2510(22) | | $5s5p^{2}(^{3}P)$ | $^{4}P_{5/2}$ | $5s5p(^{1}P^{\circ})5d$ | $^{2}D^{\circ}_{5/2}$ | 122261.3 | 366610.1 | 4.90e+08 | 0.01 | |
| | | | 409.8621(11) | | $5s5p^{2}(^{3}P)$ | $^{4}P_{1/2}$ | $5s5p(^{3}P^{\circ})5d$ | $^{2}P^{\circ}_{3/2}$ | 104226.2 | 348210.7 | 1.59e+09 | 0.04 | |
| 910bl(Cs VI) | 410.295(20) | 243727 | 410.2926(20) | 3 | $5s5p^{2}(^{1}D)$ | $^{2}D_{3/2}$ | $5s^{2}5f$ | $^{2}F^{\circ}_{5/2}$ | 141168.5 | 384897.0 | 1.49e+09 | 0.01 | TW |
| | | | 410.461(3) | | $5s^{2}5p$ | $^{2}P^{\circ}_{3/2}$ | $5s5p(^{3}P^{\circ})4f$ | $^{4}G_{5/2}$ | 19379.30 | 263007.7 | 7.47e+07 | 0.29 | |
| | | | 411.2796(23) | | $5s5p^{2}(^{3}P)$ | $^{4}P_{3/2}$ | $5s5p(^{1}P^{\circ})5d$ | $^{2}F^{\circ}_{5/2}$ | 114254.9 | 357398.5 | 3.75e+08 | 0.03 | |
| | | | 411.9232(13) | | $5s5p^{2}(^{3}P)$ | $^{4}P_{5/2}$ | $5s5p(^{1}P^{\circ})5d$ | $^{2}D^{\circ}_{3/2}$ | 122261.3 | 365025.0 | 8.64e+07 | 0.00 | |
| | | | 412.017(4) | | $5s5p^{2}(^{1}D)$ | $^{2}D_{3/2}$ | $5s5p(^{3}P^{\circ})6s$ | $^{4}P^{\circ}_{3/2}$ | 141168.5 | 383877.1 | 1.08e+09 | 0.23 | |
| | | | 413.6251(16) | | $5s5p^{2}(^{3}P)$ | $^{4}P_{3/2}$ | $5s5p(^{3}P^{\circ})5d$ | $^{2}P^{\circ}_{1/2}$ | 114254.9 | 356019.7 | 5.73e+07 | 0.01 | |
| | | | 420.298(5) | | $5s5p^{2}(^{1}D)$ | $^{2}D_{3/2}$ | $5s5p(^{3}P^{\circ})6s$ | $^{4}P^{\circ}_{1/2}$ | 141168.5 | 379095 | 7.91e+08 | 0.16 | |
| | | | 420.398(3) | | $5s5p^{2}(^{1}D)$ | $^{2}D_{5/2}$ | $5s^{2}5f$ | $^{2}F^{\circ}_{7/2}$ | 147182.2 | 385051.8 | 2.64e+08 | 0.00 | |
| | | | 422.485(4) | | $5s5p^{2}(^{1}D)$ | $^{2}D_{5/2}$ | $5s5p(^{3}P^{\circ})6s$ | $^{4}P^{\circ}_{3/2}$ | 147182.2 | 383877.1 | 4.48e+08 | 0.02 | |
| | | | 425.284(2) | | $5s5p^{2}(^{3}P)$ | $^{4}P_{5/2}$ | $5s5p(^{1}P^{\circ})5d$ | $^{2}F^{\circ}_{5/2}$ | 122261.3 | 357398.5 | 1.90e+08 | 0.01 | |
| | | | 427.4312(12) | | $5s5p^{2}(^{3}P)$ | $^{4}P_{3/2}$ | $5s5p(^{3}P^{\circ})5d$ | $^{2}P^{\circ}_{3/2}$ | 114254.9 | 348210.7 | 4.76e+08 | 0.04 | |
| 510 | 427.650(5) | 233836 | 427.650(3) | 0 | $5s5p^{2}(^{3}P)$ | $^{4}P_{5/2}$ | $5s5p(^{1}P^{\circ})5d$ | $^{2}F^{\circ}_{7/2}$ | 122261.3 | 356097.4 | 1.32e+10 | 0.08 | GJ |
| 330 | 429.356(5) | 232907 | 429.356(4) | 0 | $5s5p^{2}(^{3}P)$ | $^{2}P_{1/2}$ | $5s5p(^{3}P^{\circ})6s$ | $^{2}P^{\circ}_{1/2}$ | 158091.3 | 390998.4 | 1.30e+10 | 0.58 | GJ |
| 230 | 438.161(5) | 228227 | 438.161(3) | 0 | $5s5p^{2}(^{1}S)$ | $^{2}S_{1/2}$ | $5s5p(^{3}P^{\circ})6s$ | $^{2}P^{\circ}_{3/2}$ | 178190.9 | 406417.6 | 1.76e+10 | 0.72 | GJ |
| 90p | 438.462(5) | 228070 | 438.463(4) | -1 | $5s^{2}5d$ | $^{2}D_{3/2}$ | $5s5p(^{1}P^{\circ})6s$ | $^{2}P^{\circ}_{1/2}$ | 205994.2 | 434063.6 | 1.05e+09 | 0.11 | GJ* |
| 87 | 440.344(5) | 227095 | 440.3412(15) | 3 | $5s5p^{2}(^{1}D)$ | $^{2}D_{3/2}$ | $5s5p(^{1}P^{\circ})5d$ | $^{2}P^{\circ}_{3/2}$ | 141168.5 | 368265.1 | 3.36e+09 | 0.03 | GJ |
| 140 | 440.376(5) | 227079 | 440.377(3) | -1 | $5s5p^{2}(^{3}P)$ | $^{2}P_{3/2}$ | $5s5p(^{3}P^{\circ})6s$ | $^{2}P^{\circ}_{3/2}$ | 179339.4 | 406417.6 | 7.68e+09 | 0.30 | GJ |
| | | | 441.203(6) | | $5s^{2}5d$ | $^{2}D_{5/2}$ | $5s5p(^{1}P^{\circ})6s$ | $^{2}P^{\circ}_{3/2}$ | 208786.9 | 435440 | 4.87e+08 | 0.04 | |
| | | | 443.574(3) | | $5s5p^{2}(^{1}D)$ | $^{2}D_{3/2}$ | $5s5p(^{1}P^{\circ})5d$ | $^{2}D^{\circ}_{5/2}$ | 141168.5 | 366610.1 | 7.46e+08 | 0.00 | |
| 100 | 446.716(5) | 223856 | 446.7147(15) | 1 | $5s5p^{2}(^{1}D)$ | $^{2}D_{3/2}$ | $5s5p(^{1}P^{\circ})5d$ | $^{2}D^{\circ}_{3/2}$ | 141168.5 | 365025.0 | 3.42e+09 | 0.02 | GJ |
| 110 | 446.762(5) | 223833 | 446.7646(13) | -3 | $5s5p^{2}(^{3}P)$ | $^{4}P_{1/2}$ | $5s^{2}6p$ | $^{2}P^{\circ}_{3/2}$ | 104226.2 | 328057.7 | 7.41e+08 | 0.12 | GJ |
| | | | 446.7850(20) | | $5s5p^{2}(^{1}D)$ | $^{2}D_{3/2}$ | $5s5p(^{1}P^{\circ})5d$ | $^{2}P^{\circ}_{1/2}$ | 141168.5 | 364989.8 | 3.24e+08 | 0.00 | |
| 120 | 449.474(5) | 222482 | 449.4671(14) | 7 | $5s5p^{2}(^{3}P)$ | $^{4}P_{1/2}$ | $5s5p(^{3}P^{\circ})5d$ | $^{4}P^{\circ}_{3/2}$ | 104226.2 | 326711.9 | 1.43e+09 | 0.01 | GJ |

| $I_{Obs}$[a] (arb. u.) | $\lambda_{Obs}$[b], (Å) | $\sigma_{Obs}$, $cm^{-1}$ | $\lambda_{Ritz}$[b], (Å) | $\delta\lambda_{O-Ritz}$[c] (mÅ) | Classification | | | | $E_{low}$, $cm^{-1}$ | $E_{upp}$, $cm^{-1}$ | $gA$[d], $s^{-1}$ | CF[d] | Line Ref.[e] |
|---|---|---|---|---|---|---|---|---|---|---|---|---|---|
| | | | | | Lower level | | Upper level | | | | | | |
| 23 | 452.315(5) | 221084.9 | 452.3190(16) | -4 | $5s5p^2(^1D)$ | $^2D_{5/2}$ | $5s5p(^1P°)5d$ | $^2P°_{3/2}$ | 147182.2 | 368265.1 | 4.36e+08 | 0.00 | GJ |
| | | | 452.481(5) | | $5s5p^2(^3P)$ | $^2P_{1/2}$ | $5s5p(^3P°)6s$ | $^4P°_{1/2}$ | 158091.3 | 379095 | 8.94e+07 | 0.03 | |
| 720 | 453.845(5) | 220339.5 | 453.8439(18) | 1 | $5s5p^2(^3P)$ | $^4P_{5/2}$ | $5s5p(^3P°)5d$ | $^2F°_{7/2}$ | 122261.3 | 342601.4 | 1.43e+10 | 0.15 | GJ |
| 280 | 454.743(5) | 219904.4 | 454.7445(14) | -2 | $5s5p^2(^3P)$ | $^4P_{1/2}$ | $5s5p(^3P°)5d$ | $^2D°_{3/2}$ | 104226.2 | 324129.9 | 5.98e+09 | 0.14 | GJ |
| 180 | 458.392(5) | 218153.9 | 458.3905(15) | 1 | $5s5p^2(^3P)$ | $^4P_{3/2}$ | $5s5p(^3P°)5d$ | $^2D°_{5/2}$ | 114254.9 | 332409.5 | 2.76e+09 | 0.07 | GJ |
| 130w | 459.036(10) | 217848 | 459.0466(16) | -11 | $5s5p^2(^1D)$ | $^2D_{5/2}$ | $5s5p(^1P°)5d$ | $^2D°_{3/2}$ | 147182.2 | 365025.0 | 9.53e+08 | 0.01 | TW |
| | | | 461.6328(17) | | $5s5p^2(^3P)$ | $^4P_{1/2}$ | $5s^26p$ | $^2P°_{1/2}$ | 104226.2 | 320848.6 | 2.05e+06 | 0.01 | |
| 5100 | 462.472(5) | 216229.3 | 462.471(3) | 1 | $5s5p^2(^1D)$ | $^2D_{3/2}$ | $5s5p(^1P°)5d$ | $^2F°_{5/2}$ | 141168.5 | 357398.5 | 1.37e+11 | 0.33 | GJ |
| | | | 465.4384(20) | | $5s5p^2(^1D)$ | $^2D_{3/2}$ | $5s5p(^3P°)5d$ | $^2P°_{1/2}$ | 141168.5 | 356019.7 | 6.77e+08 | 0.01 | |
| 150 | 467.723(5) | 213801.8 | 467.7207(14) | 2 | $5s5p^2(^3P)$ | $^4P_{3/2}$ | $5s^26p$ | $^2P°_{3/2}$ | 114254.9 | 328057.7 | 4.54e+08 | 0.12 | GJ |
| | | | 469.908(5) | | $5s5p^2(^1S)$ | $^2S_{1/2}$ | $5s5p(^3P°)6s$ | $^2P°_{1/2}$ | 178190.9 | 390998.4 | 6.76e+08 | 0.04 | |
| 2700 | 470.682(5) | 212457.7 | 470.6835(16) | -1 | $5s5p^2(^3P)$ | $^4P_{3/2}$ | $5s5p(^3P°)5d$ | $^4P°_{3/2}$ | 114254.9 | 326711.9 | 4.09e+10 | 0.61 | GJ |
| 2900 | 471.470(5) | 212102.6 | 471.470(5) | | $5s5p^2(^3P)$ | $^4P_{3/2}$ | $5s5p(^3P°)5d$ | $^4P°_{1/2}$ | 114254.9 | 326357.4 | 4.38e+10 | 0.81 | GJ |
| 1800 | 472.201(5) | 211774.2 | 472.2002(22) | 1 | $5s5p^2(^3P)$ | $^4P_{3/2}$ | $5s5p(^3P°)5d$ | $^4D°_{5/2}$ | 114254.9 | 326029.5 | 3.58e+10 | 0.18 | GJ |
| | | | 472.458(5) | | $5s5p^2(^3P)$ | $^2P_{3/2}$ | $5s5p(^3P°)6s$ | $^2P°_{1/2}$ | 179339.4 | 390998.4 | 4.71e+08 | 0.05 | |
| 170 | 475.699(5) | 210217.0 | 475.701(3) | -2 | $5s5p^2(^1D)$ | $^2D_{5/2}$ | $5s5p(^1P°)5d$ | $^2F°_{5/2}$ | 147182.2 | 357398.5 | 6.37e+09 | 0.10 | GJ |
| 27 | 475.792(5) | 210175.9 | 475.7967(18) | -5 | $5s5p^2(^3P)$ | $^2P_{1/2}$ | $5s5p(^1P°)5d$ | $^2P°_{3/2}$ | 158091.3 | 368265.1 | 6.42e+09 | 0.04 | GJ |
| | | | 475.8547(17) | | $5s5p^2(^3P)$ | $^4P_{5/2}$ | $5s5p(^3P°)5d$ | $^2D°_{5/2}$ | 122261.3 | 332409.5 | 5.23e+07 | 0.00 | |
| 1000 | 476.483(5) | 209871.1 | 476.4741(16) | 9 | $5s5p^2(^3P)$ | $^4P_{3/2}$ | $5s5p(^3P°)5d$ | $^2D°_{3/2}$ | 114254.9 | 324129.9 | 2.17e+10 | 0.66 | GJ |
| 5200bl(Cs V) | 478.664(10) | 208915 | 478.663(4) | 1 | $5s5p^2(^1D)$ | $^2D_{5/2}$ | $5s5p(^1P°)5d$ | $^2F°_{7/2}$ | 147182.2 | 356097.4 | 1.77e+11 | 0.38 | GJ |
| 1600 | 479.398(5) | 208594.9 | 479.3963(14) | 2 | $5s5p^2(^3P)$ | $^4P_{3/2}$ | $5s5p(^3P°)5d$ | $^2F°_{5/2}$ | 114254.9 | 322850.6 | 2.17e+10 | 0.35 | GJ |
| 3500 | 479.846(5) | 208400.2 | 479.845(4) | 1 | $5s5p^2(^3P)$ | $^4P_{1/2}$ | $5s5p(^3P°)5d$ | $^4D°_{1/2}$ | 104226.2 | 312626.8 | 6.11e+10 | 0.80 | GJ |
| 3600 | 480.812(5) | 207981.5 | 480.8152(16) | -3 | $5s5p^2(^3P)$ | $^4P_{1/2}$ | $5s5p(^3P°)5d$ | $^4D°_{3/2}$ | 104226.2 | 312206.3 | 7.91e+10 | 0.70 | GJ |
| 130 | 482.993(5) | 207042.3 | 482.9933(15) | 0 | $5s5p^2(^1D)$ | $^2D_{3/2}$ | $5s5p(^3P°)5d$ | $^2P°_{3/2}$ | 141168.5 | 348210.7 | 3.65e+09 | 0.05 | GJ |
| 300 | 483.247(5) | 206933.5 | 483.2466(17) | 0 | $5s5p^2(^3P)$ | $^2P_{1/2}$ | $5s5p(^1P°)5d$ | $^2D°_{3/2}$ | 158091.3 | 365025.0 | 1.41e+10 | 0.06 | GJ |
| 770 | 483.330(5) | 206898.0 | 483.3288(23) | 1 | $5s5p^2(^3P)$ | $^2P_{1/2}$ | $5s5p(^1P°)5d$ | $^2P°_{1/2}$ | 158091.3 | 364989.8 | 2.81e+10 | 0.27 | GJ |
| | | | 484.0419(18) | | $5s5p^2(^3P)$ | $^4P_{3/2}$ | $5s^26p$ | $^2P°_{1/2}$ | 114254.9 | 320848.6 | 3.86e+07 | 0.21 | |
| 53000 | 485.449(5) | 205994.9 | 485.4506(13) | -2 | $5s^25p$ | $^2P°_{1/2}$ | $5s^25d$ | $^2D_{3/2}$ | 0.0 | 205994.2 | 8.82e+10 | 0.64 | GJ |
| 290 | 485.915(5) | 205797.3 | 485.9171(15) | -2 | $5s5p^2(^3P)$ | $^4P_{5/2}$ | $5s^26p$ | $^2P°_{3/2}$ | 122261.3 | 328057.7 | 2.28e+09 | 0.23 | GJ |
| | | | 486.177(5) | | $5s5p^2(^1S)$ | $^2S_{1/2}$ | $5s5p(^3P°)6s$ | $^4P°_{3/2}$ | 178190.9 | 383877.1 | 7.84e+08 | 0.12 | |
| | | | 486.482(3) | | $5s5p^2(^3P)$ | $^2P_{3/2}$ | $5s^25f$ | $^2F°_{5/2}$ | 179339.4 | 384897.0 | 4.35e+08 | 0.02 | |
| | | | 488.907(5) | | $5s5p^2(^3P)$ | $^2P_{3/2}$ | $5s5p(^3P°)6s$ | $^4P°_{3/2}$ | 179339.4 | 383877.1 | 1.02e+08 | 0.02 | |
| 1100 | 489.114(5) | 204451.3 | 489.1157(17) | -2 | $5s5p^2(^3P)$ | $^4P_{5/2}$ | $5s5p(^3P°)5d$ | $^4P°_{3/2}$ | 122261.3 | 326711.9 | 2.19e+10 | 0.44 | GJ |
| 5900 | 490.754(5) | 203768.1 | 490.7537(24) | 0 | $5s5p^2(^3P)$ | $^4P_{5/2}$ | $5s5p(^3P°)5d$ | $^4D°_{5/2}$ | 122261.3 | 326029.5 | 9.32e+10 | 0.73 | GJ |
| 25000 | 493.922(5) | 202461.1 | 493.922(3) | 0 | $5s5p^2(^3P)$ | $^4P_{5/2}$ | $5s5p(^3P°)5d$ | $^4D°_{7/2}$ | 122261.3 | 324722.6 | 2.04e+11 | 0.80 | GJ |
| 490 | 495.368(5) | 201870.1 | 495.3717(17) | -4 | $5s5p^2(^3P)$ | $^4P_{5/2}$ | $5s5p(^3P°)5d$ | $^2D°_{3/2}$ | 122261.3 | 324129.9 | 8.33e+09 | 0.47 | GJ |
| | | | 497.4419(16) | | $5s5p^2(^1D)$ | $^2D_{5/2}$ | $5s5p(^3P°)5d$ | $^2P°_{3/2}$ | 147182.2 | 348210.7 | 2.48e+08 | 0.00 | |

| $I_{Obs}$[a] (arb. u.) | $\lambda_{Obs}$[b], (Å) | $\sigma_{Obs}$, $cm^{-1}$ | $\lambda_{Ritz}$[b], (Å) | $\delta\lambda_{O-c\ Ritz}$ (mÅ) | Classification | | | | $E_{low}$, $cm^{-1}$ | $E_{upp}$, $cm^{-1}$ | $gA$[d], $s^{-1}$ | CF[d] | Line Ref.[e] |
|---|---|---|---|---|---|---|---|---|---|---|---|---|---|
| | | | | | Lower level | | Upper level | | | | | | |
| | | | 497.750(7) | | $5s5p^2(^1S)$ | $^2S_{1/2}$ | $5s5p(^3P°)6s$ | $^4P°_{1/2}$ | 178190.9 | 379095 | 8.95e+07 | 0.05 | |
| 1300 | 498.531(5) | 200589.3 | 498.5311(16) | 0 | $5s5p^2(^3P)$ | $^4P_{5/2}$ | $5s5p(^3P°)5d$ | $^2F°_{5/2}$ | 122261.3 | 322850.6 | 1.66e+10 | 0.26 | GJ |
| | | | 498.944(4) | | $5s^25d$ | $^2D_{3/2}$ | $5s5p(^3P°)6s$ | $^2P°_{3/2}$ | 205994.2 | 406417.6 | 8.76e+08 | 0.37 | |
| 5600bl(Cs VI) | 504.100(10) | 198373 | 504.104(5) | -4 | $5s5p^2(^3P)$ | $^4P_{3/2}$ | $5s5p(^3P°)5d$ | $^4D°_{1/2}$ | 114254.9 | 312626.8 | 1.52e+09 | 0.06 | GJ |
| 410 | 504.296(5) | 198296.2 | 504.2951(14) | 1 | $5s5p^2(^3P)$ | $^4P_{1/2}$ | $5p^3$ | $^2P°_{3/2}$ | 104226.2 | 302522.8 | 2.72e+09 | 0.08 | GJ |
| 1200 | 505.175(5) | 197951.2 | 505.1745(18) | 0 | $5s5p^2(^3P)$ | $^4P_{3/2}$ | $5s5p(^3P°)5d$ | $^4D°_{3/2}$ | 114254.9 | 312206.3 | 1.72e+10 | 0.21 | GJ |
| 180 | 505.235(5) | 197927.7 | 505.2332(23) | 2 | $5s5p^2(^3P)$ | $^2P_{1/2}$ | $5s5p(^3P°)5d$ | $^2P°_{1/2}$ | 158091.3 | 356019.7 | 4.74e+09 | 0.05 | GJ |
| m(Cs VI) | | | 505.994(4) | | $5s^25d$ | $^2D_{5/2}$ | $5s5p(^3P°)6s$ | $^2P°_{3/2}$ | 208786.9 | 406417.6 | 5.57e+09 | 0.53 | TW |
| 7500 | 509.068(5) | 196437.4 | 509.066(3) | 2 | $5s5p^2(^3P)$ | $^4P_{3/2}$ | $5s5p(^3P°)5d$ | $^4P°_{5/2}$ | 114254.9 | 310692.9 | 9.19e+10 | 0.74 | GJ |
| 2700 | 511.722(5) | 195418.6 | 511.7204(23) | 2 | $5s5p^2(^1D)$ | $^2D_{5/2}$ | $5s5p(^3P°)5d$ | $^2F°_{7/2}$ | 147182.2 | 342601.4 | 6.07e+10 | 0.17 | GJ |
| 9000bl(Cs V) | 522.913(10) | 191236 | 522.9004(20) | 13 | $5s5p^2(^1D)$ | $^2D_{3/2}$ | $5s5p(^3P°)5d$ | $^2D°_{5/2}$ | 141168.5 | 332409.5 | 3.35e+09 | 0.02 | GJ |
| 3300 | 525.990(5) | 190117.7 | 525.9852(18) | 5 | $5s5p^2(^3P)$ | $^2P_{1/2}$ | $5s5p(^3P°)5d$ | $^2P°_{3/2}$ | 158091.3 | 348210.7 | 9.29e+10 | 0.70 | GJ |
| 540w | 526.137(20) | 190064 | 526.1103(24) | 27 | $5s5p^2(^1S)$ | $^2S_{1/2}$ | $5s5p(^1P°)5d$ | $^2P°_{3/2}$ | 178190.9 | 368265.1 | 2.30e+09 | 0.02 | TW |
| 700 | 526.470(5) | 189944.3 | 526.4682(20) | 2 | $5s5p^2(^3P)$ | $^4P_{5/2}$ | $5s5p(^3P°)5d$ | $^4D°_{3/2}$ | 122261.3 | 312206.3 | 1.96e+09 | 0.06 | GJ |
| 36000 | 527.958(5) | 189409.0 | 527.9619(15) | -4 | $5s^25p$ | $^2P°_{3/2}$ | $5s^25d$ | $^2D_{5/2}$ | 19379.30 | 208786.9 | 1.33e+11 | 0.76 | GJ |
| | | | 528.941(6) | | $5s5p^2(^3P)$ | $^4P_{1/2}$ | $5s5p(^3P°)5d$ | $^4F°_{3/2}$ | 104226.2 | 293283.2 | 4.61e+08 | 0.18 | |
| 400p | 529.164(10) | 188977 | 529.1476(16) | 17 | $5s5p^2(^3P)$ | $^4P_{1/2}$ | $5p^3$ | $^2P°_{1/2}$ | 104226.2 | 293209.4 | 5.42e+07 | 0.01 | TW |
| 790 | 529.306(5) | 188926.6 | 529.3086(21) | -3 | $5s5p^2(^3P)$ | $^2P_{3/2}$ | $5s5p(^1P°)5d$ | $^2P°_{3/2}$ | 179339.4 | 368265.1 | 6.50e+10 | 0.38 | GJ |
| 1600 | 530.696(5) | 188431.8 | 530.697(3) | -1 | $5s5p^2(^3P)$ | $^4P_{5/2}$ | $5s5p(^3P°)5d$ | $^4P°_{5/2}$ | 122261.3 | 310692.9 | 1.63e+10 | 0.12 | GJ |
| | | | 531.1580(16) | | $5s5p^2(^3P)$ | $^4P_{3/2}$ | $5p^3$ | $^2P°_{3/2}$ | 114254.9 | 302522.8 | 1.23e+09 | 0.06 | |
| 5200bl(Cs IV) | 533.987(10) | 187270 | 533.986(4) | 1 | $5s5p^2(^3P)$ | $^2P_{3/2}$ | $5s5p(^1P°)5d$ | $^2D°_{5/2}$ | 179339.4 | 366610.1 | 1.82e+11 | 0.68 | GJ |
| 480 | 535.075(5) | 186889.7 | 535.0764(17) | -1 | $5s5p^2(^1D)$ | $^2D_{3/2}$ | $5s^26p$ | $^2P°_{3/2}$ | 141168.5 | 328057.7 | 9.77e+09 | 0.45 | GJ |
| 1700 | 535.235(5) | 186833.8 | 535.2342(22) | 1 | $5s5p^2(^1S)$ | $^2S_{1/2}$ | $5s5p(^1P°)5d$ | $^2D°_{3/2}$ | 178190.9 | 365025.0 | 8.02e+10 | 0.54 | GJ |
| | | | 535.335(3) | | $5s5p^2(^1S)$ | $^2S_{1/2}$ | $5s5p(^1P°)5d$ | $^2P°_{1/2}$ | 178190.9 | 364989.8 | 5.28e+09 | 0.07 | |
| 6200 | 535.860(5) | 186615.9 | 535.8629(14) | -3 | $5s^25p$ | $^2P°_{3/2}$ | $5s^25d$ | $^2D_{3/2}$ | 19379.30 | 205994.2 | 2.15e+10 | 0.73 | GJ |
| 580 | 538.543(5) | 185686.2 | 538.5447(21) | -2 | $5s5p^2(^3P)$ | $^2P_{3/2}$ | $5s5p(^1P°)5d$ | $^2D°_{3/2}$ | 179339.4 | 365025.0 | 1.14e+10 | 0.08 | GJ |
| 390 | 538.646(5) | 185650.7 | 538.647(3) | -1 | $5s5p^2(^3P)$ | $^2P_{3/2}$ | $5s5p(^1P°)5d$ | $^2P°_{1/2}$ | 179339.4 | 364989.8 | 1.60e+10 | 0.40 | GJ |
| 310 | 538.959(5) | 185542.9 | 538.9575(20) | 2 | $5s5p^2(^1D)$ | $^2D_{3/2}$ | $5s5p(^3P°)5d$ | $^4P°_{3/2}$ | 141168.5 | 326711.9 | 7.31e+09 | 0.36 | GJ |
| 7700bl(O II) | 539.870(10) | 185230 | 539.8772(21) | -7 | $5s5p^2(^1D)$ | $^2D_{5/2}$ | $5s5p(^3P°)5d$ | $^2D°_{5/2}$ | 147182.2 | 332409.5 | 8.01e+10 | 0.61 | GJ |
| | | | 539.989(7) | | $5s5p^2(^1D)$ | $^2D_{3/2}$ | $5s5p(^3P°)5d$ | $^4P°_{1/2}$ | 141168.5 | 326357.4 | 2.69e+08 | 0.61 | |
| | | | 540.528(6) | | $5s^25d$ | $^2D_{3/2}$ | $5s5p(^3P°)6s$ | $^2P°_{1/2}$ | 205994.2 | 390998.4 | 1.74e+09 | 0.34 | |
| 570 | 540.947(5) | 184861.0 | 540.947(3) | 0 | $5s5p^2(^1D)$ | $^2D_{3/2}$ | $5s5p(^3P°)5d$ | $^4D°_{5/2}$ | 141168.5 | 326029.5 | 8.13e+09 | 0.13 | GJ |
| 2400 | 546.560(5) | 182962.5 | 546.5634(20) | -3 | $5s5p^2(^1D)$ | $^2D_{3/2}$ | $5s5p(^3P°)5d$ | $^2D°_{3/2}$ | 141168.5 | 324129.9 | 3.71e+10 | 0.38 | GJ |
| 540 | 547.000(5) | 182815.4 | 546.998(3) | 2 | $5s5p^2(^3P)$ | $^4P_{3/2}$ | $5s5p(^3P°)5d$ | $^4F°_{5/2}$ | 114254.9 | 297070.8 | 4.45e+09 | 0.63 | GJ |
| 2400 | 550.411(5) | 181682.4 | 550.4120(18) | -1 | $5s5p^2(^1D)$ | $^2D_{3/2}$ | $5s5p(^3P°)5d$ | $^2F°_{5/2}$ | 141168.5 | 322850.6 | 4.66e+10 | 0.23 | GJ |
| | | | 552.8665(19) | | $5s5p^2(^1D)$ | $^2D_{5/2}$ | $5s^26p$ | $^2P°_{3/2}$ | 147182.2 | 328057.7 | 1.37e+09 | 0.07 | |
| 2500 | 553.719(5) | 180597.0 | 553.719(5) | | $5s5p^2(^3P)$ | $^4P_{5/2}$ | $5s5p(^3P°)5d$ | $^4F°_{7/2}$ | 122261.3 | 302858.4 | 1.57e+10 | 0.71 | GJ |

| $I_{Obs}$[a] (arb. u.) | $\lambda_{Obs}$[b], (Å) | $\sigma_{Obs}$, cm$^{-1}$ | $\lambda_{Ritz}$[b], (Å) | $\delta\lambda_{O-c}$ Ritz (mÅ) | Classification Lower level | | Upper level | | $E_{low}$, cm$^{-1}$ | $E_{upp}$, cm$^{-1}$ | $gA$[d], s$^{-1}$ | CF[d] | Line Ref.[e] |
|---|---|---|---|---|---|---|---|---|---|---|---|---|---|
| | | | 554.7496(18) | | 5s5p$^{2}$($^{3}$P) | $^{4}$P$_{5/2}$ | 5p$^{3}$ | $^{2}$P°$_{3/2}$ | 122261.3 | 302522.8 | 1.10e+08 | 0.00 | |
| 310 | 556.548(5) | 179679.0 | 556.5447(23) | 3 | 5s5p$^{2}$($^{1}$D) | $^{2}$D$_{3/2}$ | 5s$^{2}$6p | $^{2}$P°$_{1/2}$ | 141168.5 | 320848.6 | 2.82e+09 | 0.35 | GJ |
| 1900bl(Cs V) | 557.020(10) | 179527 | 557.0109(22) | 9 | 5s5p$^{2}$($^{1}$D) | $^{2}$D$_{5/2}$ | 5s5p($^{3}$P°)5d | $^{4}$P°$_{3/2}$ | 147182.2 | 326711.9 | 9.46e+09 | 0.47 | GJ |
| 14000 | 557.599(5) | 179340.4 | 557.6020(15) | -3 | 5s$^{2}$5p | $^{2}$P°$_{1/2}$ | 5s5p$^{2}$($^{3}$P) | $^{2}$P$_{3/2}$ | 0.0 | 179339.4 | 1.62e+10 | 0.30 | GJ |
| 2000bl(Cs V) | 558.584(20) | 179024 | 558.571(7) | 13 | 5s5p$^{2}$($^{3}$P) | $^{4}$P$_{3/2}$ | 5s5p($^{3}$P°)5d | $^{4}$F°$_{3/2}$ | 114254.9 | 293283.2 | 1.56e+09 | 0.59 | TW |
| 510p | 558.831(10) | 178945 | 558.8013(18) | 30 | 5s5p$^{2}$($^{3}$P) | $^{4}$P$_{3/2}$ | 5p$^{3}$ | $^{2}$P°$_{1/2}$ | 114254.9 | 293209.4 | 1.01e+08 | 0.09 | TW |
| 990 | 558.964(5) | 178902.4 | 558.963(4) | 1 | 5s$^{2}$5d | $^{2}$D$_{3/2}$ | 5s$^{2}$5f | $^{2}$F°$_{5/2}$ | 205994.2 | 384897.0 | 1.64e+11 | 0.76 | GJ |
| 230 | 559.135(5) | 178847.7 | 559.136(3) | -1 | 5s5p$^{2}$($^{1}$D) | $^{2}$D$_{5/2}$ | 5s5p($^{3}$P°)5d | $^{4}$D°$_{5/2}$ | 147182.2 | 326029.5 | 4.86e+09 | 0.14 | GJ |
| 1900 | 561.197(5) | 178190.5 | 561.1959(19) | 1 | 5s$^{2}$5p | $^{2}$P°$_{1/2}$ | 5s5p$^{2}$($^{1}$S) | $^{2}$S$_{1/2}$ | 0.0 | 178190.9 | 1.77e+09 | 0.04 | GJ |
| | | | 561.611(4) | | 5s5p$^{2}$($^{3}$P) | $^{2}$P$_{3/2}$ | 5s5p($^{1}$P°)5d | $^{2}$F°$_{5/2}$ | 179339.4 | 357398.5 | 3.88e+09 | 0.06 | |
| | | | 562.168(7) | | 5s$^{2}$5d | $^{2}$D$_{3/2}$ | 5s5p($^{3}$P°)6s | $^{4}$P°$_{3/2}$ | 205994.2 | 383877.1 | 1.47e+08 | 0.23 | |
| 2300 | 562.343(5) | 177827.4 | 562.339(3) | 4 | 5s5p$^{2}$($^{1}$S) | $^{2}$S$_{1/2}$ | 5s5p($^{3}$P°)5d | $^{2}$P°$_{1/2}$ | 178190.9 | 356019.7 | 3.53e+10 | 0.48 | GJ |
| 710 | 563.252(5) | 177540.4 | 563.252(4) | 0 | 5s5p$^{2}$($^{1}$D) | $^{2}$D$_{5/2}$ | 5s5p($^{3}$P°)5d | $^{4}$D°$_{7/2}$ | 147182.2 | 324722.6 | 1.15e+10 | 0.20 | GJ |
| 120 | 565.132(5) | 176949.8 | 565.1387(22) | -7 | 5s5p$^{2}$($^{1}$D) | $^{2}$D$_{5/2}$ | 5s5p($^{3}$P°)5d | $^{2}$D°$_{3/2}$ | 147182.2 | 324129.9 | 2.02e+09 | 0.08 | GJ |
| | | | 565.994(3) | | 5s5p$^{2}$($^{3}$P) | $^{2}$P$_{3/2}$ | 5s5p($^{3}$P°)5d | $^{2}$P°$_{1/2}$ | 179339.4 | 356019.7 | 1.17e+09 | 0.06 | |
| 1400 | 567.328(5) | 176264.9 | 567.328(5) | | 5s$^{2}$5d | $^{2}$D$_{5/2}$ | 5s$^{2}$5f | $^{2}$F°$_{7/2}$ | 208786.9 | 385051.8 | 2.19e+11 | 0.78 | GJ |
| 240 | 567.825(5) | 176110.6 | 567.827(4) | -2 | 5s$^{2}$5d | $^{2}$D$_{5/2}$ | 5s$^{2}$5f | $^{2}$F°$_{5/2}$ | 208786.9 | 384897.0 | 1.35e+10 | 0.73 | GJ |
| 650 | 569.257(5) | 175667.6 | 569.2543(20) | 3 | 5s5p$^{2}$($^{1}$D) | $^{2}$D$_{5/2}$ | 5s5p($^{3}$P°)5d | $^{2}$F°$_{5/2}$ | 147182.2 | 322850.6 | 1.01e+10 | 0.11 | GJ |
| | | | 571.134(7) | | 5s$^{2}$5d | $^{2}$D$_{5/2}$ | 5s5p($^{3}$P°)6s | $^{4}$P°$_{3/2}$ | 208786.9 | 383877.1 | 1.14e+09 | 0.40 | |
| 330 | 572.050(5) | 174809.9 | 572.051(4) | -1 | 5s5p$^{2}$($^{3}$P) | $^{4}$P$_{5/2}$ | 5s5p($^{3}$P°)5d | $^{4}$F°$_{5/2}$ | 122261.3 | 297070.8 | 3.21e+09 | 0.63 | GJ |
| | | | 577.698(9) | | 5s$^{2}$5d | $^{2}$D$_{3/2}$ | 5s5p($^{3}$P°)6s | $^{4}$P°$_{1/2}$ | 205994.2 | 379095 | 2.15e+08 | 0.34 | |
| 340 | 584.666(5) | 171037.8 | 584.6661(23) | 0 | 5s5p$^{2}$($^{1}$D) | $^{2}$D$_{3/2}$ | 5s5p($^{3}$P°)5d | $^{4}$D°$_{3/2}$ | 141168.5 | 312206.3 | 5.14e+09 | 0.33 | GJ |
| | | | 584.720(8) | | 5s5p$^{2}$($^{3}$P) | $^{4}$P$_{5/2}$ | 5s5p($^{3}$P°)5d | $^{4}$F°$_{3/2}$ | 122261.3 | 293283.2 | 7.32e+07 | 0.06 | |
| 60 | 588.155(5) | 170023.2 | 588.1668(24) | -12 | 5s5p$^{2}$($^{1}$S) | $^{2}$S$_{1/2}$ | 5s5p($^{3}$P°)5d | $^{2}$P°$_{3/2}$ | 178190.9 | 348210.7 | 2.56e+09 | 0.03 | GJ |
| | | | 588.3516(22) | | 5s5p$^{2}$($^{3}$P) | $^{2}$P$_{1/2}$ | 5s$^{2}$6p | $^{2}$P°$_{3/2}$ | 158091.3 | 328057.7 | 3.22e+08 | 0.03 | |
| | | | 589.886(4) | | 5s5p$^{2}$($^{1}$D) | $^{2}$D$_{3/2}$ | 5s5p($^{3}$P°)5d | $^{4}$P°$_{5/2}$ | 141168.5 | 310692.9 | 8.37e+08 | 0.04 | |
| 280 | 592.170(5) | 168870.4 | 592.1669(22) | 3 | 5s5p$^{2}$($^{3}$P) | $^{2}$P$_{3/2}$ | 5s5p($^{3}$P°)5d | $^{2}$P°$_{3/2}$ | 179339.4 | 348210.7 | 1.23e+10 | 0.17 | GJ |
| | | | 593.047(3) | | 5s5p$^{2}$($^{3}$P) | $^{2}$P$_{1/2}$ | 5s5p($^{3}$P°)5d | $^{4}$P°$_{3/2}$ | 158091.3 | 326711.9 | 2.41e+09 | 0.10 | |
| 1500 | 600.083(5) | 166643.6 | 600.0816(16) | 1 | 5s5p$^{2}$($^{3}$P) | $^{4}$P$_{1/2}$ | 5p$^{3}$ | $^{4}$S°$_{3/2}$ | 104226.2 | 270870.2 | 6.09e+09 | 0.43 | GJ |
| 770 | 602.267(5) | 166039.3 | 602.2696(24) | -3 | 5s5p$^{2}$($^{3}$P) | $^{2}$P$_{1/2}$ | 5s5p($^{3}$P°)5d | $^{2}$D°$_{3/2}$ | 158091.3 | 324129.9 | 9.32e+09 | 0.13 | GJ |
| | | | 605.972(3) | | 5s5p$^{2}$($^{1}$D) | $^{2}$D$_{5/2}$ | 5s5p($^{3}$P°)5d | $^{4}$D°$_{3/2}$ | 147182.2 | 312206.3 | 1.71e+08 | 0.01 | |
| 520 | 611.580(5) | 163510.9 | 611.581(4) | -1 | 5s5p$^{2}$($^{1}$D) | $^{2}$D$_{5/2}$ | 5s5p($^{3}$P°)5d | $^{4}$P°$_{5/2}$ | 147182.2 | 310692.9 | 6.35e+09 | 0.11 | GJ |
| 490 | 616.257(5) | 162270.0 | 616.253(3) | 4 | 5s$^{2}$5d | $^{2}$D$_{3/2}$ | 5s5p($^{1}$P°)5d | $^{2}$P°$_{3/2}$ | 205994.2 | 368265.1 | 2.15e+10 | 0.45 | GJ |
| 5100bl(O IV) | 616.948(10) | 162088 | 616.948(10) | | 5s$^{2}$6s | $^{2}$S$_{1/2}$ | 5s5p($^{1}$P°)6s | $^{2}$P°$_{3/2}$ | 273351.5 | 435440 | 4.95e+10 | 0.59 | GJ* |
| 120 | 619.757(5) | 161353.6 | 619.7542(20) | 3 | 5s5p$^{2}$($^{1}$D) | $^{2}$D$_{3/2}$ | 5p$^{3}$ | $^{2}$P°$_{3/2}$ | 141168.5 | 302522.8 | 2.37e+09 | 0.04 | GJ |
| 1100p | 622.233(5) | 160711.5 | 622.231(5) | 2 | 5s$^{2}$6s | $^{2}$S$_{1/2}$ | 5s5p($^{1}$P°)6s | $^{2}$P°$_{1/2}$ | 273351.5 | 434063.6 | 2.39e+10 | 0.59 | GJ* |
| | | | 622.603(5) | | 5s$^{2}$5d | $^{2}$D$_{3/2}$ | 5s5p($^{1}$P°)5d | $^{2}$D°$_{5/2}$ | 205994.2 | 366610.1 | 1.02e+08 | 0.00 | |

| $I_{Obs}$[a] (arb. u.) | $\lambda_{Obs}$[b], (Å) | $\sigma_{Obs}$, cm$^{-1}$ | $\lambda_{Ritz}$[b], (Å) | $\delta\lambda_{O-Ritz}$[c] (mÅ) | Classification: Lower level | | Classification: Upper level | | $E_{low}$, cm$^{-1}$ | $E_{upp}$, cm$^{-1}$ | $gA$[d], s$^{-1}$ | CF[d] | Line Ref.[e] |
|---|---|---|---|---|---|---|---|---|---|---|---|---|---|
| 4700bl(O IV) | 625.146(10) | 159963 | 625.1559(15) | -10 | $5s^25p$ | $^2P°_{3/2}$ | $5s5p^2(^3P)$ | $^2P_{3/2}$ | 19379.30 | 179339.4 | 5.95e+10 | 0.55 | GJ |
| 700 | 627.045(5) | 159478.2 | 627.045(3) | 0 | $5s^25d$ | $^2D_{5/2}$ | $5s5p(^1P°)5d$ | $^2P°_{3/2}$ | 208786.9 | 368265.1 | 2.97e+10 | 0.28 | GJ |
| 470 | 628.810(5) | 159030.5 | 628.809(3) | 1 | $5s^25d$ | $^2D_{3/2}$ | $5s5p(^1P°)5d$ | $^2D°_{3/2}$ | 205994.2 | 365025.0 | 1.77e+10 | 0.22 | GJ |
| 450 | 628.948(5) | 158995.7 | 628.948(4) | 0 | $5s^25d$ | $^2D_{3/2}$ | $5s5p(^1P°)5d$ | $^2P°_{1/2}$ | 205994.2 | 364989.8 | 1.97e+10 | 0.33 | GJ |
| 9500 | 629.679(5) | 158811.1 | 629.6769(21) | 2 | $5s^25p$ | $^2P°_{3/2}$ | $5s5p^2(^1S)$ | $^2S_{1/2}$ | 19379.30 | 178190.9 | 2.01e+10 | 0.59 | GJ |
| 190 | 630.435(5) | 158620.6 | 630.436(4) | -1 | $5s^24f$ | $^2F°_{5/2}$ | $5s5p(^3P°)4f$ | $^2D_{5/2}$ | 166536.93 | 325157.4 | 3.85e+09 | 0.38 | GJ |
| | | | 632.3271(21) | | $5s5p^2(^3P)$ | $^4P_{3/2}$ | $5p^3$ | $^2D°_{5/2}$ | 114254.9 | 272400.9 | 7.16e+07 | 0.05 | |
| 8200 | 632.544(5) | 158091.8 | 632.5459(19) | -2 | $5s^25p$ | $^2P°_{1/2}$ | $5s5p^2(^3P)$ | $^2P_{1/2}$ | 0.0 | 158091.3 | 2.70e+10 | 0.65 | GJ |
| 3800 | 633.470(5) | 157860.7 | 633.469(4) | 1 | $5s^24f$ | $^2F°_{7/2}$ | $5s5p(^3P°)4f$ | $^2D_{5/2}$ | 167296.5 | 325157.4 | 7.57e+10 | 0.67 | GJ |
| 720 | 633.620(5) | 157823.3 | 633.620(5) | 0 | $5s^25d$ | $^2D_{5/2}$ | $5s5p(^1P°)5d$ | $^2D°_{5/2}$ | 208786.9 | 366610.1 | 4.88e+10 | 0.42 | GJ |
| 7400 | 635.337(5) | 157396.8 | 635.337(5) | | $5s^24f$ | $^2F°_{5/2}$ | $5s5p(^3P°)4f$ | $^2D_{3/2}$ | 166536.93 | 323933.7 | 5.24e+10 | 0.69 | GJ |
| 7600 | 638.508(5) | 156615.1 | 638.5072(18) | 1 | $5s5p^2(^3P)$ | $^4P_{3/2}$ | $5p^3$ | $^4S°_{3/2}$ | 114254.9 | 270870.2 | 1.48e+10 | 0.64 | GJ |
| 2200 | 639.341(5) | 156411.1 | 639.3384(18) | 3 | $5s5p^2(^3P)$ | $^4P_{1/2}$ | $5p^3$ | $^2D°_{3/2}$ | 104226.2 | 260637.9 | 7.31e+09 | 0.49 | GJ |
| 84 | 640.049(5) | 156238.0 | 640.049(3) | 0 | $5s^25d$ | $^2D_{5/2}$ | $5s5p(^1P°)5d$ | $^2D°_{3/2}$ | 208786.9 | 365025.0 | 1.25e+10 | 0.29 | GJ |
| | | | 642.359(7) | | $5s5p^2(^1D)$ | $^2D_{5/2}$ | $5s5p(^3P°)5d$ | $^4F°_{7/2}$ | 147182.2 | 302858.4 | 2.04e+07 | 0.00 | |
| 1200 | 643.747(5) | 155340.5 | 643.7467(22) | 0 | $5s5p^2(^1D)$ | $^2D_{5/2}$ | $5p^3$ | $^2P°_{3/2}$ | 147182.2 | 302522.8 | 1.48e+10 | 0.27 | GJ |
| | | | 647.101(8) | | $5s5p^2(^3P)$ | $^2P_{1/2}$ | $5s5p(^3P°)5d$ | $^4D°_{1/2}$ | 158091.3 | 312626.8 | 1.35e+08 | 0.02 | |
| 190 | 648.857(10) | 154117.2 | 648.866(3) | -9 | $5s5p^2(^3P)$ | $^2P_{1/2}$ | $5s5p(^3P°)5d$ | $^4D°_{3/2}$ | 158091.3 | 312206.3 | 3.50e+08 | 0.02 | TW |
| 75 | 653.295(5) | 153070.2 | 653.295(3) | 0 | $5s5p^2(^3P)$ | $^2P_{3/2}$ | $5s5p(^3P°)5d$ | $^2D°_{5/2}$ | 179339.4 | 332409.5 | 1.36e+09 | 0.01 | GJ |
| 860 | 657.396(10) | 152115.3 | 657.399(9) | -3 | $5s5p^2(^1D)$ | $^2D_{3/2}$ | $5s5p(^3P°)5d$ | $^4F°_{3/2}$ | 141168.5 | 293283.2 | 1.23e+09 | 0.27 | TW |
| 1100 | 657.713(5) | 152042.0 | 657.7178(23) | -5 | $5s5p^2(^1D)$ | $^2D_{3/2}$ | $5p^3$ | $^2P°_{1/2}$ | 141168.5 | 293209.4 | 1.21e+10 | 0.37 | GJ |
| 1500bl(Cs IV) | 660.483(10) | 151404.4 | 660.483(6) | 0 | $5s^25d$ | $^2D_{3/2}$ | $5s5p(^1P°)5d$ | $^2F°_{5/2}$ | 205994.2 | 357398.5 | 2.64e+10 | 0.15 | GJ |
| 3400 | 661.419(5) | 151190.1 | 661.420(4) | -1 | $5s^24f$ | $^2F°_{5/2}$ | $5s5p(^3P°)4f$ | $^2G_{7/2}$ | 166536.93 | 317726.7 | 6.30e+10 | 0.51 | GJ |
| 560 | 664.761(5) | 150430.0 | 664.760(4) | 1 | $5s^24f$ | $^2F°_{7/2}$ | $5s5p(^3P°)4f$ | $^2G_{7/2}$ | 167296.5 | 317726.7 | 7.93e+09 | 0.63 | GJ |
| 520 | 666.046(5) | 150139.8 | 666.0468(23) | -1 | $5s5p^2(^3P)$ | $^4P_{5/2}$ | $5p^3$ | $^2D°_{5/2}$ | 122261.3 | 272400.9 | 2.75e+09 | 0.18 | GJ |
| 150 | 666.549(5) | 150026.5 | 666.553(4) | -4 | $5s^25d$ | $^2D_{3/2}$ | $5s5p(^3P°)5d$ | $^2P°_{1/2}$ | 205994.2 | 356019.7 | 1.12e+09 | 0.08 | GJ |
| 5100 | 666.874(5) | 149953.4 | 666.874(5) | | $5s^24f$ | $^2F°_{7/2}$ | $5s5p(^3P°)4f$ | $^2G_{9/2}$ | 167296.5 | 317249.8 | 9.31e+10 | 0.58 | GJ |
| | | | 667.162(5) | | $5s5p^2(^1D)$ | $^2D_{5/2}$ | $5s5p(^3P°)5d$ | $^4F°_{5/2}$ | 147182.2 | 297070.8 | 1.02e+09 | 0.21 | |
| | | | 667.259(3) | | $5s5p^2(^1S)$ | $^2S_{1/2}$ | $5s^26p$ | $^2P°_{3/2}$ | 178190.9 | 328057.7 | 1.62e+08 | 0.03 | |
| | | | 672.412(3) | | $5s5p^2(^3P)$ | $^2P_{3/2}$ | $5s^26p$ | $^2P°_{3/2}$ | 179339.4 | 328057.7 | 1.18e+09 | 0.16 | |
| | | | 672.895(6) | | $5s^25d$ | $^2D_{5/2}$ | $5s5p(^1P°)5d$ | $^2F°_{5/2}$ | 208786.9 | 357398.5 | 1.07e+08 | 0.00 | |
| 2700bl(O II) | 672.926(10) | 148604.8 | 672.9072(22) | 19 | $5s5p^2(^3P)$ | $^4P_{5/2}$ | $5p^3$ | $^4S°_{3/2}$ | 122261.3 | 270870.2 | 1.11e+10 | 0.44 | GJ |
| | | | 673.305(4) | | $5s5p^2(^1S)$ | $^2S_{1/2}$ | $5s5p(^3P°)5d$ | $^4P°_{3/2}$ | 178190.9 | 326711.9 | 7.14e+07 | 0.01 | |
| | | | 678.553(3) | | $5s5p^2(^3P)$ | $^2P_{3/2}$ | $5s5p(^3P°)5d$ | $^4P°_{3/2}$ | 179339.4 | 326711.9 | 1.46e+08 | 0.02 | |
| 1600* | 678.837(14) | 147311 | 678.831(5) | 6 | $5s^24f$ | $^2F°_{5/2}$ | $5s5p(^3P°)4f$ | $^2F_{7/2}$ | 166536.93 | 313848.9 | 9.23e+09 | 0.66 | GJ |
| 1600* | 678.837(14) | 147311 | 678.838(9) | -1 | $5s^25d$ | $^2D_{5/2}$ | $5s5p(^1P°)5d$ | $^2F°_{7/2}$ | 208786.9 | 356097.4 | 3.32e+10 | 0.18 | GJ |
| | | | 680.189(11) | | $5s5p^2(^3P)$ | $^2P_{3/2}$ | $5s5p(^3P°)5d$ | $^4P°_{1/2}$ | 179339.4 | 326357.4 | 8.81e+07 | 0.49 | |

| $I_{Obs}$[a] (arb. u.) | $\lambda_{Obs}$[b], (Å) | $\sigma_{Obs}$, $cm^{-1}$ | $\lambda_{Ritz}$[b], (Å) | $\delta\lambda_{O-Ritz}$[c] (mÅ) | Classification | | | | $E_{low}$, $cm^{-1}$ | $E_{upp}$, $cm^{-1}$ | $gA$[d], $s^{-1}$ | CF[d] | Line Ref.[e] |
|---|---|---|---|---|---|---|---|---|---|---|---|---|---|
| | | | | | Lower level | | Upper level | | | | | | |
| | | | 681.709(5) | | $5s5p^2(^3P)$ | $^2P_{3/2}$ | $5s5p(^3P°)5d$ | $^4D°_{5/2}$ | 179339.4 | 326029.5 | 1.70e+08 | 0.03 | |
| 3000 | 681.982(5) | 146631.4 | 681.982(4) | 0 | $5s^24f$ | $^2F°_{5/2}$ | $5s5p(^3P°)4f$ | $^2F_{5/2}$ | 166536.93 | 313168.4 | 5.26e+10 | 0.65 | GJ |
| 6400 | 682.349(5) | 146552.6 | 682.350(5) | -1 | $5s^24f$ | $^2F°_{7/2}$ | $5s5p(^3P°)4f$ | $^2F_{7/2}$ | 167296.5 | 313848.9 | 6.72e+10 | 0.64 | GJ |
| 1600 | 683.137(5) | 146383.5 | 683.1394(21) | -2 | $5s5p^2(^3P)$ | $^4P_{3/2}$ | $5p^3$ | $^2D°_{3/2}$ | 114254.9 | 260637.9 | 7.09e+09 | 0.42 | GJ |
| | | | 684.458(10) | | $5s5p^2(^1D)$ | $^2D_{5/2}$ | $5s5p(^3P°)5d$ | $^4F°_{3/2}$ | 147182.2 | 293283.2 | 5.63e+08 | 0.40 | |
| 58 | 685.225(5) | 145937.5 | 685.218(3) | 7 | $5s5p^2(^1S)$ | $^2S_{1/2}$ | $5s5p(^3P°)5d$ | $^2D°_{3/2}$ | 178190.9 | 324129.9 | 9.69e+08 | 0.03 | GJ |
| 150 | 685.533(5) | 145871.9 | 685.533(4) | 0 | $5s^24f$ | $^2F°_{7/2}$ | $5s5p(^3P°)4f$ | $^2F_{5/2}$ | 167296.5 | 313168.4 | 2.44e+09 | 0.13 | GJ |
| | | | 690.653(3) | | $5s5p^2(^3P)$ | $^2P_{3/2}$ | $5s5p(^3P°)5d$ | $^2D°_{3/2}$ | 179339.4 | 324129.9 | 6.26e+08 | 0.02 | |
| 200 | 692.370(5) | 144431.4 | 692.3697(24) | 0 | $5s5p^2(^3P)$ | $^2P_{1/2}$ | $5p^3$ | $^2P°_{3/2}$ | 158091.3 | 302522.8 | 2.23e+09 | 0.04 | GJ |
| 65 | 696.810(5) | 143511.1 | 696.810(3) | 0 | $5s5p^2(^3P)$ | $^2P_{3/2}$ | $5s5p(^3P°)5d$ | $^2F°_{5/2}$ | 179339.4 | 322850.6 | 1.36e+09 | 0.02 | GJ |
| 120 | 703.155(5) | 142216.2 | 703.153(3) | 2 | $5s^25d$ | $^2D_{3/2}$ | $5s5p(^3P°)5d$ | $^2P°_{3/2}$ | 205994.2 | 348210.7 | 4.36e+09 | 0.16 | GJ |
| | | | 706.668(4) | | $5s5p^2(^3P)$ | $^2P_{3/2}$ | $5s^26p$ | $^2P°_{1/2}$ | 179339.4 | 320848.6 | 1.79e+08 | 0.13 | |
| 4600 | 708.370(5) | 141169.2 | 708.3733(23) | -3 | $5s^25p$ | $^2P°_{1/2}$ | $5s5p^2(^1D)$ | $^2D_{3/2}$ | 0.0 | 141168.5 | 9.06e+09 | 0.21 | GJ |
| 59 | 717.239(5) | 139423.5 | 717.238(3) | 1 | $5s^25d$ | $^2D_{5/2}$ | $5s5p(^3P°)5d$ | $^2P°_{3/2}$ | 208786.9 | 348210.7 | 3.16e+09 | 0.23 | GJ |
| 910 | 720.916(5) | 138712.4 | 720.9182(21) | -2 | $5s^25p$ | $^2P°_{3/2}$ | $5s5p^2(^3P)$ | $^2P_{1/2}$ | 19379.30 | 158091.3 | 9.11e+08 | 0.04 | GJ |
| 2200 | 722.661(5) | 138377.5 | 722.6655(24) | -5 | $5s5p^2(^3P)$ | $^4P_{5/2}$ | $5p^3$ | $^2D°_{3/2}$ | 122261.3 | 260637.9 | 1.50e+10 | 0.62 | GJ |
| | | | 737.126(23) | | $5s^24f$ | $^2F°_{5/2}$ | $5s5p(^1P°)4f$ | $^2D_{3/2}$ | 166536.93 | 302199 | 2.51e+08 | 0.01 | |
| | | | 739.689(12) | | $5s5p^2(^3P)$ | $^2P_{1/2}$ | $5s5p(^3P°)5d$ | $^4F°_{3/2}$ | 158091.3 | 293283.2 | 2.96e+07 | 0.01 | |
| 43 | 740.087(5) | 135119.2 | 740.093(3) | -6 | $5s5p^2(^3P)$ | $^2P_{1/2}$ | $5p^3$ | $^2P°_{1/2}$ | 158091.3 | 293209.4 | 6.01e+08 | 0.02 | GJ |
| | | | 746.183(4) | | $5s5p^2(^1S)$ | $^2S_{1/2}$ | $5s5p(^3P°)5d$ | $^4D°_{3/2}$ | 178190.9 | 312206.3 | 4.56e+07 | 0.00 | |
| 69 | 747.302(5) | 133814.7 | 747.303(4) | -1 | $5s^25d$ | $^2D_{5/2}$ | $5s5p(^3P°)5d$ | $^2F°_{7/2}$ | 208786.9 | 342601.4 | 6.10e+09 | 0.08 | GJ |
| | | | 751.506(14) | | $5s^26s$ | $^2S_{1/2}$ | $5s5p(^3P°)6s$ | $^2P°_{3/2}$ | 273351.5 | 406417.6 | 7.20e+08 | 0.44 | |
| 35 | 752.635(5) | 132866.5 | 752.633(4) | 2 | $5s5p^2(^3P)$ | $^2P_{3/2}$ | $5s5p(^3P°)5d$ | $^4D°_{3/2}$ | 179339.4 | 312206.3 | 3.76e+08 | 0.04 | GJ |
| | | | 761.304(6) | | $5s5p^2(^3P)$ | $^2P_{3/2}$ | $5s5p(^3P°)5d$ | $^4P°_{5/2}$ | 179339.4 | 310692.9 | 2.47e+08 | 0.01 | |
| 230bl(O V) | 761.995(20) | 131234 | 762.007(3) | -12 | $5s5p^2(^1D)$ | $^2D_{3/2}$ | $5p^3$ | $^2D°_{5/2}$ | 141168.5 | 272400.9 | 2.11e+09 | 0.11 | TW |
| 150 | 770.996(5) | 129702.4 | 771.000(2) | -4 | $5s5p^2(^1D)$ | $^2D_{3/2}$ | $5p^3$ | $^4S°_{3/2}$ | 141168.5 | 270870.2 | 2.07e+09 | 0.09 | GJ |
| 210 | 773.203(5) | 129332.1 | 773.205(4) | -2 | $5s^24f$ | $^2F°_{5/2}$ | $5s5p(^1P°)4f$ | $^2D_{5/2}$ | 166536.93 | 295868.7 | 2.99e+09 | 0.17 | GJ |
| 54 | 777.775(5) | 128571.9 | 777.773(4) | 2 | $5s^24f$ | $^2F°_{7/2}$ | $5s5p(^1P°)4f$ | $^2D_{5/2}$ | 167296.5 | 295868.7 | 7.99e+08 | 0.02 | GJ |
| 6900 | 782.451(5) | 127803.5 | 782.455(2) | -4 | $5s^25p$ | $^2P°_{3/2}$ | $5s5p^2(^1D)$ | $^2D_{5/2}$ | 19379.30 | 147182.2 | 6.08e+09 | 0.15 | GJ |
| | | | 791.043(5) | | $5s^25d$ | $^2D_{3/2}$ | $5s5p(^3P°)5d$ | $^2D°_{5/2}$ | 205994.2 | 332409.5 | 2.94e+08 | 0.01 | |
| 1100 | 798.601(5) | 125219.0 | 798.603(3) | -2 | $5s5p^2(^1D)$ | $^2D_{5/2}$ | $5p^3$ | $^2D°_{5/2}$ | 147182.2 | 272400.9 | 1.02e+10 | 0.24 | GJ |
| 43 | 804.300(5) | 124331.7 | 804.299(3) | 1 | $5s5p^2(^1S)$ | $^2S_{1/2}$ | $5p^3$ | $^2P°_{3/2}$ | 178190.9 | 302522.8 | 1.02e+09 | 0.03 | GJ |
| 380 | 808.486(5) | 123688.0 | 808.486(3) | 0 | $5s5p^2(^1D)$ | $^2D_{5/2}$ | $5p^3$ | $^4S°_{3/2}$ | 147182.2 | 270870.2 | 3.49e+09 | 0.45 | GJ |
| 41 | 808.913(5) | 123622.7 | 808.914(4) | -1 | $5s^25d$ | $^2D_{5/2}$ | $5s5p(^3P°)5d$ | $^2D°_{5/2}$ | 208786.9 | 332409.5 | 1.79e+09 | 0.08 | GJ |
| 810 | 811.795(5) | 123183.8 | 811.798(3) | -3 | $5s5p^2(^3P)$ | $^2P_{3/2}$ | $5p^3$ | $^2P°_{3/2}$ | 179339.4 | 302522.8 | 6.97e+09 | 0.14 | GJ |
| 66 | 813.447(5) | 122933.6 | 813.447(4) | 0 | $5s^24f$ | $^2F°_{5/2}$ | $5s5p(^1P°)4f$ | $^2G_{7/2}$ | 166536.93 | 289470.5 | 9.80e+08 | 0.03 | GJ |
| 200 | 818.505(5) | 122174.0 | 818.504(4) | 1 | $5s^24f$ | $^2F°_{7/2}$ | $5s5p(^1P°)4f$ | $^2G_{7/2}$ | 167296.5 | 289470.5 | 2.06e+09 | 0.09 | GJ |

| $I_{Obs}$[a] (arb. u.) | $\lambda_{Obs}$[b], (Å) | $\sigma_{Obs}$, $cm^{-1}$ | $\lambda_{Ritz}$[b], (Å) | $\delta\lambda_{O-C\ Ritz}$ (mÅ) | Classification: Lower level | | Upper level | | $E_{low}$, $cm^{-1}$ | $E_{upp}$, $cm^{-1}$ | $gA$[d], $s^{-1}$ | CF[d] | Line Ref.[e] |
|---|---|---|---|---|---|---|---|---|---|---|---|---|---|
| 27 | 819.245(5) | 122063.6 | 819.246(4) | -1 | $5s^25d$ | $^2D_{3/2}$ | $5s^26p$ | $^2P°_{3/2}$ | 205994.2 | 328057.7 | 1.54e+09 | 0.24 | GJ |
| 180 | 821.094(5) | 121788.7 | 821.091(3) | 3 | $5s^25p$ | $^2P°_{3/2}$ | $5s5p^2(^1D)$ | $^2D_{3/2}$ | 19379.30 | 141168.5 | 1.49e+08 | 0.01 | GJ |
| | | | 828.379(5) | | $5s^25d$ | $^2D_{3/2}$ | $5s5p(^3P°)5d$ | $^4P°_{3/2}$ | 205994.2 | 326711.9 | 6.82e+07 | 0.03 | |
| 28 | 831.455(5) | 120271.1 | 831.453(5) | 2 | $5s^24f$ | $^2F°_{5/2}$ | $5s5p(^3P°)4f$ | $^4D_{5/2}$ | 166536.93 | 286808.3 | 4.14e+08 | 0.35 | GJ |
| | | | 833.088(7) | | $5s^25d$ | $^2D_{3/2}$ | $5s5p(^3P°)5d$ | $^4D°_{5/2}$ | 205994.2 | 326029.5 | 2.17e+08 | 0.03 | |
| 350 | 837.038(5) | 119468.9 | 837.034(3) | 4 | $5s5p^2(^1D)$ | $^2D_{3/2}$ | $5p^3$ | $^2D°_{3/2}$ | 141168.5 | 260637.9 | 2.17e+09 | 0.08 | GJ |
| 390 | 838.430(5) | 119270.5 | 838.428(4) | 2 | $5s^25d$ | $^2D_{5/2}$ | $5s^26p$ | $^2P°_{3/2}$ | 208786.9 | 328057.7 | 1.08e+10 | 0.58 | GJ |
| 22 | 847.994(5) | 117925.4 | 847.997(4) | -3 | $5s^25d$ | $^2D_{5/2}$ | $5s5p(^3P°)5d$ | $^4P°_{3/2}$ | 208786.9 | 326711.9 | 3.76e+08 | 0.10 | GJ |
| 9 | 848.187(5) | 117898.5 | 848.187(4) | 0 | $5s^24f$ | $^2F°_{5/2}$ | $5s5p(^3P°)4f$ | $4F_{7/2}$ | 166536.93 | 284435.5 | 1.80e+08 | 0.04 | GJ |
| | | | 852.932(8) | | $5s^25d$ | $^2D_{5/2}$ | $5s5p(^3P°)5d$ | $^4D°_{5/2}$ | 208786.9 | 326029.5 | 4.02e+08 | 0.21 | |
| 48 | 853.686(5) | 117139.1 | 853.686(4) | 0 | $5s^24f$ | $^2F°_{7/2}$ | $5s5p(^3P°)4f$ | $4F_{7/2}$ | 167296.5 | 284435.5 | 1.04e+09 | 0.46 | GJ |
| 31 | 855.750(5) | 116856.6 | 855.751(4) | -1 | $5s^25d$ | $^2D_{3/2}$ | $5s5p(^3P°)5d$ | $^2F°_{5/2}$ | 205994.2 | 322850.6 | 1.79e+09 | 0.07 | GJ |
| | | | 862.547(10) | | $5s^25d$ | $^2D_{5/2}$ | $5s5p(^3P°)5d$ | $^4D°_{7/2}$ | 208786.9 | 324722.6 | 2.44e+08 | 0.13 | |
| | | | 866.979(6) | | $5s^25d$ | $^2D_{5/2}$ | $5s5p(^3P°)5d$ | $^2D°_{3/2}$ | 208786.9 | 324129.9 | 1.52e+08 | 0.02 | |
| 180bl(Cs VI/2) | 869.424(10) | 115018.7 | 869.425(5) | -1 | $5s5p^2(^1S)$ | $^2S_{1/2}$ | $5p^3$ | $^2P°_{1/2}$ | 178190.9 | 293209.4 | 3.41e+09 | 0.17 | GJ |
| 160 | 870.666(5) | 114854.6 | 870.668(5) | -2 | $5s^25d$ | $^2D_{3/2}$ | $5s^26p$ | $^2P°_{1/2}$ | 205994.2 | 320848.6 | 5.76e+09 | 0.61 | GJ |
| 18 | 875.240(5) | 114254.4 | 875.236(3) | 4 | $5s^25p$ | $^2P°_{1/2}$ | $5s5p^2(^3P)$ | $^4P_{3/2}$ | 0.0 | 114254.9 | 8.57e+06 | 0.02 | GJ |
| | | | 876.703(5) | | $5s^25d$ | $^2D_{5/2}$ | $5s5p(^3P°)5d$ | $^2F°_{5/2}$ | 208786.9 | 322850.6 | 1.35e+08 | 0.02 | |
| 25 | 878.197(5) | 113869.7 | 878.194(4) | 3 | $5s5p^2(^3P)$ | $^2P_{3/2}$ | $5p^3$ | $^2P°_{1/2}$ | 179339.4 | 293209.4 | 3.50e+08 | 0.04 | GJ |
| 46 | 881.404(5) | 113455.4 | 881.401(3) | 3 | $5s5p^2(^1D)$ | $^2D_{5/2}$ | $5p^3$ | $^2D°_{3/2}$ | 147182.2 | 260637.9 | 7.18e+08 | 0.05 | GJ |
| 170 | 886.696(5) | 112778.2 | 886.691(3) | 5 | $5s5p^2(^3P)$ | $^2P_{1/2}$ | $5p^3$ | $^4S°_{3/2}$ | 158091.3 | 270870.2 | 1.71e+09 | 0.15 | GJ |
| | | | 904.768(23) | | $5s^26s$ | $^2S_{1/2}$ | $5s5p(^3P°)6s$ | $^4P°_{3/2}$ | 273351.5 | 383877.1 | 8.98e+08 | 0.49 | |
| 18 | 905.507(5) | 110435.4 | 905.508(4) | -1 | $5s^24f$ | $^2F°_{5/2}$ | $5s5p(^1P°)4f$ | $^2F_{7/2}$ | 166536.93 | 276972.2 | 4.69e+08 | 0.05 | GJ |
| 39 | 911.780(5) | 109675.6 | 911.779(4) | 1 | $5s^24f$ | $^2F°_{7/2}$ | $5s5p(^1P°)4f$ | $^2F_{7/2}$ | 167296.5 | 276972.2 | 3.56e+08 | 0.03 | GJ |
| 18 | 922.398(5) | 108413.1 | 922.392(4) | 6 | $5s^24f$ | $^2F°_{5/2}$ | $5s5p(^1P°)4f$ | $^2F_{5/2}$ | 166536.93 | 274950.7 | 4.21e+07 | 0.00 | GJ |
| 14 | 928.894(5) | 107654.9 | 928.900(4) | -6 | $5s^24f$ | $^2F°_{7/2}$ | $5s5p(^1P°)4f$ | $^2F_{5/2}$ | 167296.5 | 274950.7 | 1.61e+08 | 0.03 | GJ |
| | | | 937.800(17) | | $5s^25d$ | $^2D_{3/2}$ | $5s5p(^3P°)5d$ | $^4D°_{1/2}$ | 205994.2 | 312626.8 | 1.66e+08 | 0.18 | |
| | | | 941.512(7) | | $5s^25d$ | $^2D_{3/2}$ | $5s5p(^3P°)5d$ | $^4D°_{3/2}$ | 205994.2 | 312206.3 | 1.67e+08 | 0.07 | |
| | | | 955.122(10) | | $5s^25d$ | $^2D_{3/2}$ | $5s5p(^3P°)5d$ | $^4P°_{5/2}$ | 205994.2 | 310692.9 | 1.01e+08 | 0.07 | |
| 590 | 959.452(5) | 104226.2 | 959.452(4) | 0 | $5s^25p$ | $^2P°_{1/2}$ | $5s5p^2(^3P)$ | $^4P_{1/2}$ | 0.0 | 104226.2 | 2.84e+08 | 0.59 | GJ |
| | | | 966.937(7) | | $5s^25d$ | $^2D_{5/2}$ | $5s5p(^3P°)5d$ | $^4D°_{3/2}$ | 208786.9 | 312206.3 | 9.30e+07 | 0.05 | |
| 1200 | 971.984(5) | 102882.4 | 971.987(4) | -3 | $5s^25p$ | $^2P°_{3/2}$ | $5s5p^2(^3P)$ | $^4P_{5/2}$ | 19379.30 | 122261.3 | 8.13e+08 | 0.23 | GJ |
| 140 | 975.163(5) | 102547.0 | 975.166(4) | -3 | $5s5p^2(^3P)$ | $^2P_{1/2}$ | $5p^3$ | $^2D°_{3/2}$ | 158091.3 | 260637.9 | 1.90e+09 | 0.11 | GJ |
| 55 | 975.744(5) | 102485.9 | 975.745(4) | -1 | $5s^24f$ | $^2F°_{5/2}$ | $5s5p(^3P°)4f$ | $^4D_{7/2}$ | 166536.93 | 269022.7 | 5.96e+08 | 0.25 | GJ |
| 83 | 979.304(5) | 102113.3 | 979.307(4) | -3 | $5s^24f$ | $^2F°_{5/2}$ | $5s5p(^3P°)4f$ | $4F_{5/2}$ | 166536.93 | 268650.0 | 5.97e+08 | 0.10 | GJ |
| | | | 981.296(10) | | $5s^25d$ | $^2D_{5/2}$ | $5s5p(^3P°)5d$ | $^4P°_{5/2}$ | 208786.9 | 310692.9 | 3.18e+08 | 0.07 | |
| 96 | 983.031(5) | 101726.2 | 983.031(4) | 0 | $5s^24f$ | $^2F°_{7/2}$ | $5s5p(^3P°)4f$ | $^4D_{7/2}$ | 167296.5 | 269022.7 | 8.74e+08 | 0.05 | GJ |

| $I_{Obs}$[a] (arb. u.) | $\lambda_{Obs}$[b], (Å) | $\sigma_{Obs}$, $cm^{-1}$ | $\lambda_{Ritz}$[b], (Å) | $\delta\lambda_{O-c\ Ritz}$ (mÅ) | Classification | | | | $E_{low}$, $cm^{-1}$ | $E_{upp}$, $cm^{-1}$ | $gA$[d], $s^{-1}$ | CF[d] | Line Ref.[e] |
|---|---|---|---|---|---|---|---|---|---|---|---|---|---|
| | | | | | Lower level | | Upper level | | | | | | |
| 26 | 983.965(5) | 101629.6 | 983.965(5) | 0 | $5s^24f$ | $^2F°_{5/2}$ | $5s5p(^3P°)4f$ | $4F_{3/2}$ | 166536.93 | 268166.6 | 3.70e+08 | 0.33 | GJ |
| 36 | 984.383(5) | 101586.5 | 984.383(5) | | $5s^24f$ | $^2F°_{7/2}$ | $5s5p(^3P°)4f$ | $^4G_{9/2}$ | 167296.5 | 268882.9 | 7.88e+08 | 0.31 | GJ |
| 20 | 986.647(5) | 101353.4 | 986.645(4) | 2 | $5s^24f$ | $^2F°_{7/2}$ | $5s5p(^3P°)4f$ | $4F_{5/2}$ | 167296.5 | 268650.0 | 2.71e+08 | 0.15 | GJ |
| 62 | 1013.300(10) | 98687.4 | 1013.300(10) | | $5s^24f$ | $^2F°_{5/2}$ | $5s5p(^3P°)4f$ | $^4G_{7/2}$ | 166536.93 | 265224.4 | 2.79e+08 | 0.37 | TW |
| | | | 1035.962(7) | | $5s^25d$ | $^2D_{3/2}$ | $5p^3$ | $^2P°_{3/2}$ | 205994.2 | 302522.8 | 8.03e+07 | 0.03 | |
| 47 | 1036.583(10) | 96470.8 | 1036.583(10) | | $5s^24f$ | $^2F°_{5/2}$ | $5s5p(^3P°)4f$ | $^4G_{5/2}$ | 166536.93 | 263007.7 | 2.05e+08 | 0.07 | TW |
| | | | 1044.809(11) | | $5s^24f$ | $^2F°_{7/2}$ | $5s5p(^3P°)4f$ | $^4G_{5/2}$ | 167296.5 | 263007.7 | 2.08e+07 | 0.04 | |
| 130 | 1054.015(5) | 94875.3 | 1054.012(4) | 3 | $5s^25p$ | $^2P°_{3/2}$ | $5s5p^2(^3P)$ | $^4P_{3/2}$ | 19379.30 | 114254.9 | 8.65e+07 | 0.42 | GJ |
| | | | 1063.021(20) | | $5s^25d$ | $^2D_{5/2}$ | $5s5p(^3P°)5d$ | $^4F°_{7/2}$ | 208786.9 | 302858.4 | 4.94e+08 | 0.34 | |
| 230 | 1074.560(5) | 93061.3 | 1074.558(4) | 2 | $5s5p^2(^3P)$ | $^2P_{3/2}$ | $5p^3$ | $^2D°_{5/2}$ | 179339.4 | 272400.9 | 3.08e+09 | 0.12 | GJ |
| | | | 1078.990(7) | | $5s5p^2(^1S)$ | $^2S_{1/2}$ | $5p^3$ | $^4S°_{3/2}$ | 178190.9 | 270870.2 | 1.42e+08 | 0.04 | |
| 12 | 1092.525(5) | 91531.1 | 1092.528(4) | -3 | $5s5p^2(^3P)$ | $^2P_{3/2}$ | $5p^3$ | $^4S°_{3/2}$ | 179339.4 | 270870.2 | 2.82e+08 | 0.07 | GJ |
| | | | 1097.977(15) | | $5s^25d$ | $^2D_{3/2}$ | $5s5p(^3P°)5d$ | $^4F°_{5/2}$ | 205994.2 | 297070.8 | 2.39e+08 | 0.26 | |
| | | | 1121.29(3) | | $5s5p(^1P°)4f$ | $^2D_{5/2}$ | $5s^25f$ | $^2F°_{7/2}$ | 295868.7 | 385051.8 | 2.35e+08 | 0.26 | |
| | | | 1145.62(3) | | $5s^25d$ | $^2D_{3/2}$ | $5s5p(^3P°)5d$ | $^4F°_{3/2}$ | 205994.2 | 293283.2 | 4.71e+07 | 0.08 | |
| 26 | 1178.596(5) | 84846.7 | 1178.593(4) | 3 | $5s^25p$ | $^2P°_{3/2}$ | $5s5p^2(^3P)$ | $^4P_{1/2}$ | 19379.30 | 104226.2 | 2.90e+07 | 0.09 | GJ |
| | | | 1209.22(6) | | $5s5p(^1P°)4f$ | $^2D_{3/2}$ | $5s^25f$ | $^2F°_{5/2}$ | 302199 | 384897.0 | 2.66e+08 | 0.29 | |
| | | | 1212.900(9) | | $5s5p^2(^1S)$ | $^2S_{1/2}$ | $5p^3$ | $^2D°_{3/2}$ | 178190.9 | 260637.9 | 1.00e+08 | 0.01 | |
| | | | 1230.035(7) | | $5s5p^2(^3P)$ | $^2P_{3/2}$ | $5p^3$ | $^2D°_{3/2}$ | 179339.4 | 260637.9 | 4.32e+07 | 0.00 | |
| | | | 1571.981(15) | | $5s^25d$ | $^2D_{5/2}$ | $5p^3$ | $^2D°_{5/2}$ | 208786.9 | 272400.9 | 2.35e+07 | 0.02 | |
| | | | 1790.84(7) | | $5s5p(^3P°)4f$ | $^4G_{9/2}$ | $5s5p(^3P°)5d$ | $^4D°_{7/2}$ | 268882.9 | 324722.6 | 1.30e+08 | 0.06 | |
| | | | 1827.95(6) | | $5s^26s$ | $^2S_{1/2}$ | $5s^26p$ | $^2P°_{3/2}$ | 273351.5 | 328057.7 | 2.87e+09 | 0.63 | |
| | | | 1874.05(7) | | $5s^26s$ | $^2S_{1/2}$ | $5s5p(^3P°)5d$ | $^4P°_{3/2}$ | 273351.5 | 326711.9 | 1.92e+08 | 0.61 | |
| | | | 1910.51(6) | | $5s5p(^1P°)4f$ | $^2D_{5/2}$ | $5s5p(^3P°)5d$ | $^2P°_{3/2}$ | 295868.7 | 348210.7 | 1.83e+08 | 0.11 | |
| | | | 1912.70(6) | | $5s5p^2(^3P)$ | $^4P_{3/2}$ | $5s^24f$ | $^2F°_{5/2}$ | 114254.9 | 166536.93 | 1.09e+06 | 0.51 | |
| | | | 1969.34(8) | | $5s^26s$ | $^2S_{1/2}$ | $5s5p(^3P°)5d$ | $^2D°_{3/2}$ | 273351.5 | 324129.9 | 1.13e+08 | 0.58 | |
| | | | 2032.73(7) | | $5s5p(^1P°)4f$ | $^2F_{5/2}$ | $5s5p(^3P°)5d$ | $^2D°_{3/2}$ | 274950.7 | 324129.9 | 1.32e+08 | 0.10 | |
| | | | 2104.72(9) | | $5s^26s$ | $^2S_{1/2}$ | $5s^26p$ | $^2P°_{1/2}$ | 273351.5 | 320848.6 | 1.08e+09 | 0.65 | |
| | | | 2178.99(8) | | $5s5p(^1P°)4f$ | $^2F_{7/2}$ | $5s5p(^3P°)5d$ | $^2F°_{5/2}$ | 276972.2 | 322850.6 | 1.96e+08 | 0.13 | |
| | | | 2219.80(8) | | $5s5p^2(^3P)$ | $^4P_{5/2}$ | $5s^24f$ | $^2F°_{7/2}$ | 122261.3 | 167296.5 | 1.32e+07 | 0.50 | |
| | | | 2248.51(12) | | $5s5p(^3P°)4f$ | $^4F_{3/2}$ | $5s5p(^3P°)5d$ | $^4D°_{1/2}$ | 268166.6 | 312626.8 | 1.31e+08 | 0.32 | |
| | | | 2295.17(9) | | $5s5p(^3P°)4f$ | $^4F_{5/2}$ | $5s5p(^3P°)5d$ | $^4D°_{3/2}$ | 268650.0 | 312206.3 | 1.80e+08 | 0.30 | |
| | | | 2328.17(9) | | $5s5p(^1P°)4f$ | $^2G_{7/2}$ | $5s5p(^3P°)5d$ | $^2D°_{5/2}$ | 289470.5 | 332409.5 | 2.13e+08 | 0.16 | |
| | | | 2366.14(9) | | $5s^24f$ | $^2F°_{5/2}$ | $5s^25d$ | $^2D_{5/2}$ | 166536.93 | 208786.9 | 1.25e+07 | 0.21 | |
| | | | 2399.07(10) | | $5s5p(^3P°)4f$ | $^4D_{7/2}$ | $5s5p(^3P°)5d$ | $^4P°_{5/2}$ | 269022.7 | 310692.9 | 2.11e+08 | 0.28 | |
| | | | 2403.46(11) | | $5s5p(^3P°)4f$ | $^4F_{7/2}$ | $5s5p(^3P°)5d$ | $^4D°_{5/2}$ | 284435.5 | 326029.5 | 1.88e+08 | 0.20 | |
| | | | 2409.46(9) | | $5s^24f$ | $^2F°_{7/2}$ | $5s^25d$ | $^2D_{5/2}$ | 167296.5 | 208786.9 | 2.64e+08 | 0.25 | |

| $I_{Obs}$ [a] (arb. u.) | $\lambda_{Obs}$ [b], (Å) | $\sigma_{Obs}$, cm$^{-1}$ | $\lambda_{Ritz}$ [b], (Å) | $\delta\lambda_{O-Ritz}$ [c] (mÅ) | Classification | | | | $E_{low}$, cm$^{-1}$ | $E_{upp}$, cm$^{-1}$ | $gA$ [d], s$^{-1}$ | CF [d] | Line Ref. [e] |
|---|---|---|---|---|---|---|---|---|---|---|---|---|---|
| | | | | | Lower level | | Upper level | | | | | | |
| | | | 2533.63(10) | | $5s^24f$ | $^2F°_{5/2}$ | $5s^25d$ | $^2D_{3/2}$ | 166536.93 | 205994.2 | 1.51e+08 | 0.24 | |
| | | | 2942.44(20) | | $5s5p(^3P°)4f$ | $^4G_{9/2}$ | $5s5p(^3P°)5d$ | $^4F°_{7/2}$ | 268882.9 | 302858.4 | 2.15e+08 | 0.46 | |
| | | | 3139.16(21) | | $5s5p(^3P°)4f$ | $^4G_{7/2}$ | $5s5p(^3P°)5d$ | $^4F°_{5/2}$ | 265224.4 | 297070.8 | 1.39e+08 | 0.48 | |
| | | | 3302.1(3) | | $5s5p(^3P°)4f$ | $^4G_{5/2}$ | $5s5p(^3P°)5d$ | $^4F°_{3/2}$ | 263007.7 | 293283.2 | 9.63e+07 | 0.49 | |
| | | | 3940.79(24) | | $5s5p^2(^1D)$ | $^2D_{3/2}$ | $5s^24f$ | $^2F°_{5/2}$ | 141168.5 | 166536.93 | 1.30e+07 | 0.51 | |
| | | | 4970.2(4) | | $5s5p^2(^1D)$ | $^2D_{5/2}$ | $5s^24f$ | $^2F°_{7/2}$ | 147182.2 | 167296.5 | 8.44e+06 | 0.48 | |

[a] Observed relative intensities, in terms of total energy flux under the line profile, are reduced to a common arbitrary linear scale corresponding to a plasma in local thermodynamic equilibrium with an effective excitation temperature of 6.6 eV, which we derived for the exposures of Gayasov and Joshi [9] (see section 3.4). Line character of the observed line: bl—blended by a close line (the blending spectrum is indicated in parentheses); w—wide line; *—intensity shared by two or more transitions; m—masked by a stronger neighboring line (no wavelength measured); p—intensity of the line is perturbed.

[b] Observed and Ritz wavelengths (in Å) are given in vacuum for wavenumbers ($\sigma$) ≥ 50000 cm$^{-1}$, and in standard air smaller wavenumbers. The quantity given in parentheses is the uncertainty in the last digit.

[c] Difference between the observed and Ritz wavelengths in milliangstrom, 1 mÅ =10$^{-3}$ Å.

[d] Weighted transition probability (*gA*-value) and cancelation factor from the HFR calculations (see the text).

[e] Reference to the observed line: GJ–Gayasov and Joshi [9]; GJ*–reported by Gayasov and Joshi [9] with our additional line character; TW–this work. Blank for the predicted lines whose Ritz wavelengths are computed from the energy levels involved.

Table 2. Optimized energy levels of Cs VII.

| Configuration | Level | Energy,[a] ($cm^{-1}$) | | Unc[b] | Leading percentages[c] | | | | | | | $\Delta E_{(O-C)}$[d] | No. of Lines[e] |
|---|---|---|---|---|---|---|---|---|---|---|---|---|---|
| | | | | | %1 | %2 | Conf. | Term | %3 | Conf. | Term | | |
| $5s^25p$ | $^2P^\circ_{1/2}$ | 0.0 | | 0.5 | 99 | | | | | | | 151 | 9 |
| $5s^25p$ | $^2P^\circ_{3/2}$ | 19379.3 | | 0.4 | 98 | | | | | | | -150 | 16 |
| $5s5p^2(^3P)$ | $^4P_{1/2}$ | 104226.2 | | 0.4 | 93 | 6 | $5s5p^2(^1S)$ | $^2S$ | | | | 61 | 13 |
| $5s5p^2(^3P)$ | $^4P_{3/2}$ | 114254.9 | | 0.4 | 98 | | | | | | | -104 | 19 |
| $5s5p^2(^3P)$ | $^4P_{5/2}$ | 122261.3 | | 0.5 | 84 | 15 | $5s5p^2(^1D)$ | $^2D$ | | | | 33 | 17 |
| $5s5p^2(^1D)$ | $^2D_{3/2}$ | 141168.5 | B1 | - | 85 | 7 | $5s^25d$ | $^2D$ | 5 | $5s5p^2(^3P)$ | $^2P$ | -105 | 22 |
| $5s5p^2(^1D)$ | $^2D_{5/2}$ | 147182.2 | | 0.4 | 76 | 16 | $5s5p^2(^3P)$ | $^4P$ | 7 | $5s^25d$ | $^2D$ | 94 | 19 |
| $5s5p^2(^3P)$ | $^2P_{1/2}$ | 158091.3 | | 0.4 | 65 | 29 | $5s5p^2(^1S)$ | $^2S$ | 5 | $5s5p^2(^3P)$ | $^4P$ | 292 | 14 |
| $5s^24f$ | $^2F^\circ_{5/2}$ | 166536.93 | B2 | - | 98 | | | | | | | 219 | 13 |
| $5s^24f$ | $^2F^\circ_{7/2}$ | 167296.5 | | 0.3 | 98 | | | | | | | -219 | 11 |
| $5s5p^2(^1S)$ | $^2S_{1/2}$ | 178190.9 | | 0.6 | 64 | 32 | $5s5p^2(^3P)$ | $^2P$ | | | | -126 | 10 |
| $5s5p^2(^3P)$ | $^2P_{3/2}$ | 179339.4 | | 0.4 | 91 | 4 | $5s5p^2(^1D)$ | $^2D$ | | | | -153 | 16 |
| $5s^25d$ | $^2D_{3/2}$ | 205994.2 | | 0.5 | 87 | 8 | $5s5p^2(^1D)$ | $^2D$ | | | | 14 | 13 |
| $5s^25d$ | $^2D_{5/2}$ | 208786.9 | | 0.5 | 89 | 7 | $5s5p^2(^1D)$ | $^2D$ | | | | -6 | 11 |
| $5p^3$ | $^2D^\circ_{3/2}$ | 260637.9 | | 0.4 | 38 | 30 | $5p^3$ | $^4S^\circ$ | 22 | $5p^3$ | $^2P^\circ$ | 150 | 6 |
| $5s5p(^3P^\circ)4f$ | $^4G_{5/2}$ | 263007.7 | N | 0.9 | 72 | 16 | $5s5p(^3P^\circ)4f$ | $^4F$ | 6 | $5s5p(^3P^\circ)4f$ | $^2F$ | 123 | 1 |
| $5s5p(^3P^\circ)4f$ | $^4G_{7/2}$ | 265224.4 | N | 1.0 | 55 | 37 | $5s5p(^3P^\circ)4f$ | $^4F$ | | | | 146 | 1 |
| $5s5p(^3P^\circ)4f$ | $^4F_{3/2}$ | 268166.6 | | 0.5 | 89 | 6 | $5s5p(^3P^\circ)4f$ | $^4D$ | | | | -154 | 2 |
| $5s5p(^3P^\circ)4f$ | $^4F_{5/2}$ | 268650.0 | | 0.4 | 56 | 16 | $5s5p(^3P^\circ)4f$ | $^4D$ | 11 | $5s5p(^3P^\circ)4f$ | $^2F$ | -373 | 3 |
| $5s5p(^3P^\circ)4f$ | $^4G_{9/2}$ | 268882.9 | | 0.6 | 47 | 47 | $5s5p(^3P^\circ)4f$ | $^4F$ | 5 | $5s5p(^3P^\circ)4f$ | $^2G$ | 27 | 1 |
| $5s5p(^3P^\circ)4f$ | $^4D_{7/2}$ | 269022.7 | | 0.4 | 27 | 24 | $5s5p(^3P^\circ)4f$ | $^2F$ | 20 | $5s5p(^1P^\circ)4f$ | $^2F$ | -87 | 2 |
| $5p^3$ | $^4S^\circ_{3/2}$ | 270870.2 | | 0.4 | 51 | 38 | $5p^3$ | $^2D^\circ$ | 10 | $5s5p(^3P^\circ)5d$ | $^2D^\circ$ | -133 | 7 |
| $5p^3$ | $^2D^\circ_{5/2}$ | 272400.9 | | 0.5 | 79 | 19 | $5s5p(^3P^\circ)5d$ | $^2D^\circ$ | | | | -21 | 4 |
| $5s^26s$ | $^2S_{1/2}$ | 273351.5 | | 1.9 | 98 | | | | | | | 0 | 3 |
| $5s5p(^1P^\circ)4f$ | $^2F_{5/2}$ | 274950.7 | | 0.4 | 50 | 18 | $5s5p(^3P^\circ)4f$ | $^4G$ | 12 | $5s5p(^1P^\circ)4f$ | $^2D$ | -137 | 3 |
| $5s5p(^1P^\circ)4f$ | $^2F_{7/2}$ | 276972.2 | | 0.5 | 29 | 21 | $5s5p(^3P^\circ)4f$ | $^2G$ | 20 | $5s5p(^1P^\circ)4f$ | $^4G$ | -183 | 2 |
| $5s5p(^3P^\circ)4f$ | $^4F_{9/2}$ | (281300) | L | 300 | 50 | 50 | $5s5p(^3P^\circ)4f$ | $^4G$ | | | | | |
| $5s5p(^3P^\circ)4f$ | $^4G_{11/2}$ | (282500) | L | 300 | 100 | | | | | | | | |
| $5s5p(^3P^\circ)4f$ | $^4F_{7/2}$ | 284435.5 | | 0.5 | 45 | 30 | $5s5p(^3P^\circ)4f$ | $^4D$ | 12 | $5s5p(^1P^\circ)4f$ | $^2G$ | 338 | 2 |
| $5s5p(^3P^\circ)4f$ | $^4D_{5/2}$ | 286808.3 | | 0.7 | 71 | 21 | $5s5p(^3P^\circ)4f$ | $^4F$ | | | | 195 | 2 |
| $5s5p(^3P^\circ)4f$ | $^4D_{3/2}$ | (288700) | L | 300 | 91 | 8 | $5s5p(^3P^\circ)4f$ | $^4F$ | | | | | |
| $5s5p(^1P^\circ)4f$ | $^2G_{7/2}$ | 289470.5 | | 0.6 | 52 | 15 | $5s5p(^3P^\circ)4f$ | $^4D$ | 15 | $5s5p(^3P^\circ)4f$ | $^2F$ | -393 | 2 |
| $5s5p(^3P^\circ)4f$ | $^4D_{1/2}$ | (289900) | L | 300 | 100 | | | | | | | | |
| $5p^3$ | $^2P^\circ_{1/2}$ | 293209.4 | | 0.5 | 88 | 8 | $5s5p(^3P^\circ)5d$ | $^2P^\circ$ | | | | -96 | 6 |
| $5s5p(^3P^\circ)5d$ | $^4F^\circ_{3/2}$ | 293283.2 | N | 2.2 | 92 | 2 | $5s5p(^3P^\circ)5d$ | $^2D^\circ$ | | | | -257 | 2 |
| $5s5p(^1P^\circ)4f$ | $^2D_{5/2}$ | 295868.7 | | 0.6 | 59 | 16 | $5s5p(^3P^\circ)4f$ | $^2F$ | 11 | $5s5p(^3P^\circ)4f$ | $^2D$ | 452 | 3 |

| Configuration | Level | Energy,[a] (cm$^{-1}$) | | Unc[b] | Leading percentages[c] | | | | | | | $\Delta E_{(O-C)}$[d] | No. of Lines[e] |
|---|---|---|---|---|---|---|---|---|---|---|---|---|---|
| | | | | | %1 | %2 | Conf. | Term | %3 | Conf. | Term | | |
| 5s5p($^3$P°)5d | $^4$F°$_{5/2}$ | 297070.8 | | 1.2 | 91 | 4 | 5s5p($^3$P°)5d | $^4$D° | | | | -174 | 2 |
| 5s5p($^1$P°)4f | $^2$G$_{9/2}$ | (298300) | L | 300 | 76 | 21 | 5s5p($^3$P°)4f | $^2$G | | | | | |
| 5s5p($^1$P°)4f | $^2$D$_{3/2}$ | 302199 | | 4 | 65 | 29 | 5s5p($^3$P°)4f | $^2$D | | | | 12 | 1 |
| 5p$^3$ | $^2$P°$_{3/2}$ | 302522.8 | | 0.5 | 59 | 11 | 5s5p($^3$P°)5d | $^2$P° | 10 | 5p$^3$ | $^4$S° | 99 | 6 |
| 5s5p($^3$P°)5d | $^4$F°$_{7/2}$ | 302858.4 | | 1.7 | 89 | 8 | 5s5p($^3$P°)5d | $^4$D° | | | | -231 | 1 |
| 5s5p($^3$P°)5d | $^4$P°$_{5/2}$ | 310692.9 | | 1.0 | 56 | 25 | 5s5p($^3$P°)5d | $^4$D° | 9 | 5s5p($^3$P°)5d | $^2$D° | 683 | 3 |
| 5s5p($^3$P°)5d | $^4$D°$_{3/2}$ | 312206.3 | | 0.7 | 58 | 32 | 5s5p($^3$P°)5d | $^4$P° | | | | 99 | 6 |
| 5s5p($^3$P°)5d | $^4$D°$_{1/2}$ | 312626.8 | | 1.9 | 85 | 9 | 5s5p($^3$P°)5d | $^4$P° | | | | -426 | 2 |
| 5s5p($^3$P°)4f | $^2$F$_{5/2}$ | 313168.4 | | 0.8 | 55 | 33 | 5s5p($^1$P°)4f | $^2$F | 7 | 5s5p($^1$P°)4f | $^2$D | 515 | 2 |
| 5s5p($^3$P°)5d | $^4$F°$_{9/2}$ | (313500) | L | 400 | 99 | | | | | | | | |
| 5s5p($^3$P°)4f | $^2$F$_{7/2}$ | 313848.9 | | 1.1 | 53 | 38 | 5s5p($^1$P°)4f | $^2$F | 5 | 5s5p($^1$P°)4f | $^2$G | 60 | 2 |
| 5s5p($^3$P°)4f | $^2$G$_{9/2}$ | 317249.8 | | 1.2 | 73 | 23 | 5s5p($^1$P°)4f | $^2$G | | | | -36 | 1 |
| 5s5p($^3$P°)4f | $^2$G$_{7/2}$ | 317726.7 | | 0.8 | 90 | 6 | 5s5p($^1$P°)4f | $^2$G | | | | -168 | 2 |
| 5s$^2$6p | $^2$P°$_{1/2}$ | 320848.6 | | 0.8 | 97 | | | | | | | -114 | 2 |
| 5s5p($^3$P°)5d | $^2$F°$_{5/2}$ | 322850.6 | | 0.6 | 30 | 22 | 5s5p($^3$P°)5d | $^2$D° | 16 | 5s5p($^1$P°)5d | $^2$F° | -57 | 6 |
| 5s5p($^3$P°)4f | $^2$D$_{3/2}$ | 323933.7 | | 1.2 | 62 | 33 | 5s5p($^1$P°)4f | $^2$D | | | | -134 | 1 |
| 5s5p($^3$P°)5d | $^2$D°$_{3/2}$ | 324129.9 | | 0.7 | 48 | 13 | 5s5p($^3$P°)5d | $^4$D° | 10 | 5p$^3$ | $^2$D° | -343 | 7 |
| 5s5p($^3$P°)5d | $^4$D°$_{7/2}$ | 324722.6 | | 1.3 | 88 | 9 | 5s5p($^3$P°)5d | $^4$F° | | | | -50 | 2 |
| 5s5p($^3$P°)4f | $^2$D$_{5/2}$ | 325157.4 | | 0.9 | 78 | 17 | 5s5p($^1$P°)4f | $^2$D | | | | -207 | 3 |
| 5s5p($^3$P°)5d | $^4$D°$_{5/2}$ | 326029.5 | | 1.0 | 53 | 26 | 5s5p($^3$P°)5d | $^4$P° | 13 | 5s5p($^3$P°)5d | $^2$F° | -10 | 4 |
| 5s5p($^3$P°)5d | $^4$P°$_{1/2}$ | 326357.4 | | 2.3 | 90 | 10 | 5s5p($^3$P°)5d | $^4$D° | | | | 357 | 1 |
| 5s5p($^3$P°)5d | $^4$P°$_{3/2}$ | 326711.9 | | 0.7 | 58 | 24 | 5s5p($^3$P°)5d | $^4$D° | 7 | 5s5p($^3$P°)5d | $^2$D° | 176 | 6 |
| 5s$^2$6p | $^2$P°$_{3/2}$ | 328057.7 | | 0.6 | 86 | 5 | 5s5p($^3$P°)5d | $^2$D° | | | | 128 | 6 |
| 5s5p($^3$P°)5d | $^2$D°$_{5/2}$ | 332409.5 | | 0.7 | 32 | 23 | 5s5p($^3$P°)5d | $^2$F° | 16 | 5s5p($^3$P°)5d | $^4$P° | 523 | 5 |
| 5s5p($^3$P°)5d | $^2$F°$_{7/2}$ | 342601.4 | | 0.9 | 64 | 32 | 5s5p($^1$P°)5d | $^2$F° | | | | -63 | 3 |
| 5s5p($^3$P°)5d | $^2$P°$_{3/2}$ | 348210.7 | | 0.6 | 55 | 28 | 5s5p($^1$P°)5d | $^2$D° | 9 | 5p$^3$ | $^2$P° | 403 | 6 |
| 5s5p($^3$P°)5d | $^2$P°$_{1/2}$ | 356019.7 | | 0.9 | 80 | 8 | 5s5p($^1$P°)5d | $^2$P° | 5 | 5p$^3$ | $^2$P° | -497 | 3 |
| 5s5p($^1$P°)5d | $^2$F°$_{7/2}$ | 356097.4 | | 1.9 | 52 | 30 | 5s5p($^3$P°)5d | $^2$F° | 12 | 5s$^2$5f | $^2$F° | 61 | 3 |
| 5s5p($^1$P°)5d | $^2$F°$_{5/2}$ | 357398.5 | | 1.3 | 55 | 21 | 5s5p($^3$P°)5d | $^2$F° | 14 | 5s$^2$5f | $^2$F° | -190 | 3 |
| 5s5p($^1$P°)5d | $^2$P°$_{1/2}$ | 364989.8 | | 1.0 | 82 | 7 | 5s5p($^3$P°)5d | $^2$P° | 5 | 5p$^3$ | $^2$P° | -474 | 3 |
| 5s5p($^1$P°)5d | $^2$D°$_{3/2}$ | 365025.0 | | 0.7 | 47 | 22 | 5s5p($^3$P°)5d | $^2$P° | 12 | 5s5p($^1$P°)5d | $^2$P° | 173 | 7 |
| 5s5p($^1$P°)5d | $^2$D°$_{5/2}$ | 366610.1 | | 1.3 | 67 | 12 | 5s5p($^3$P°)5d | $^2$D° | 6 | 5s5p($^3$P°)5d | $^2$F° | 67 | 2 |
| 5s5p($^1$P°)5d | $^2$P°$_{3/2}$ | 368265.1 | | 0.8 | 77 | 6 | 5s5p($^1$P°)5d | $^2$D° | 5 | 5s5p($^3$P°)5d | $^2$D° | 190 | 7 |
| 5s5p($^3$P°)6s | $^4$P°$_{1/2}$ | 379095 | | 3 | 94 | 4 | 5s5p($^3$P°)6s | $^2$P° | | | | -26 | 2 |
| 5s5p($^3$P°)6s | $^4$P°$_{3/2}$ | 383877.1 | | 2.1 | 88 | 7 | 5s5p($^3$P°)6s | $^2$P° | | | | -8 | 3 |
| 5s$^2$5f | $^2$F°$_{5/2}$ | 384897.0 | | 1.2 | 83 | 12 | 5s5p($^1$P°)5d | $^2$F° | | | | -105 | 3 |
| 5s$^2$5f | $^2$F°$_{7/2}$ | 385051.8 | | 1.6 | 85 | 11 | 5s5p($^1$P°)5d | $^2$F° | | | | 124 | 1 |
| 5s5p($^3$P°)6s | $^2$P°$_{1/2}$ | 390998.4 | | 2.1 | 91 | 4 | 5s5p($^3$P°)6s | $^4$P° | | | | -71 | 2 |

| Configuration | Level | Energy,[a] (cm$^{-1}$) | Unc[b] | Leading percentages[c] | | | | | | | ΔE$_{(O-C)}$[d] | No. of Lines[e] |
|---|---|---|---|---|---|---|---|---|---|---|---|---|
| | | | | %1 | %2 | Conf. | Term | %3 | Conf. | Term | | |
| 5s5p($^3$P°)6s | $^4$P°$_{5/2}$ | 398620.2 | 2.1 | 99 | | | | | | | -1 | 3 |
| 5s5p($^3$P°)6s | $^2$P°$_{3/2}$ | 406417.6 | 1.7 | 88 | 9 | 5s5p($^3$P°)6s | $^4$P° | | | | 111 | 3 |
| 5s5p($^1$P°)6s | $^2$P°$_{1/2}$ | 434063.6 | 1.9 | 81 | 14 | 5s$^2$7p | $^2$P° | | | | 15 | 3 |
| 5s5p($^1$P°)6s | $^2$P°$_{3/2}$ | 435440 | 3 | 85 | 7 | 5s$^2$7p | $^2$P° | | | | -17 | 1 |

[a] Optimized energy values. The auxiliary symbols inserted next to energy values have the following meaning: B1 or B2–the base levels chosen for representing the level uncertainties given in next column (see the next footnote); L–The given level values and their uncertainties are the theoretical ones from the LSF of Cowan's code (see section 3.1); N–newly identified energy levels.

[b] Uncertainties of the level values given in the preceding column. They are given on the level of one standard deviation. They correspond to uncertainties of level separations from 5s5p$^2$($^1$D) $^2D_{3/2}$ (base level B1) for all levels except those of the 5s$^2$4f and 5s5p4f configurations, for which the uncertainties are given for separations from 5s$^2$4f $^2$F°$_{5/2}$, which is designated as the base level B2. The uncertainties of the base levels B1 and B2 relative to the ground level are 0.5 cm$^{-1}$ and 1.5 cm$^{-1}$, respectively. To determine the absolute uncertainties of excitation energies with respect to the ground level, the given values should be combined in quadrature the uncertainty of the corresponding base level.

[c] The *LS*-coupling percentage compositions were determined in this work by a parametric least-squares fitting with Cowan's codes. The first percentage value refers to the configuration and term given in the first two columns of the table. The remaining percentage values refer to the configurations and terms given next to them.

[d] Differences between observed and calculated energies from the parametric least squares fitting (see text). Blank for unobserved levels.

[e] Number of observed lines determining the level in the optimization procedure. Blank for experimentally unknown levels.

### 3.1. HFR calculations and transition probabilities

We use the pseudo-relativistic Hartree-Fock (HFR) approach with superposition of interacting configurations implemented in Cowan's suite codes [13] to obtain the theoretical support for our analysis on Cs VII. The odd-parity matrix contains $5s^2np$ ($n$=5–9), $5s^2nf$ ($n$=4–7), $5p^3$, 5s5p(5d+6d+6s+7s), 5s5d4f, $5p^2np$ ($n$=6–9), and $5p^2nf$ ($n$=4–7) configurations; whereas the $5s5p^2$, $5s^2n\ell$ ($n$=6–9 for $\ell$=s, $n$=5–9 for $\ell$=d, and $n$=5–7 for $\ell$=g), $5s5d^2$, 5s5p(4f+5f+6p+7p), 4f5s6p, $5p^2n\ell$ ($n$=6–9 for $\ell$=s, $n$=5–9 for $\ell$=d, and $n$=5–7 for $\ell$=g), and $5s4f^2$ configurations were included in the even-parity system. In the initial calculations the Slater parameters, the $F^k$ was fixed at 85%, ($G^k$ and $R^k$) were fixed at 75%, and ($E_{av}$ and $\zeta_{n\ell}$) at 100% of their HFR values. A least-squares-parametric fitting (LSF) was performed to minimize the differences between observed and theoretical energy values (see **Table 3**). In the odd parity set 42 known levels were fitted with 16 free parameters, and the fits converged with a standard deviation (SD) of 303 $cm^{-1}$, whereas 30 known levels of even parity were fitted with 14 free parameters, and an SD of 284 $cm^{-1}$ was achieved. Using these fitted energy parameters, the transition probabilities (TPs or $gA$-values) were re-calculated and are given in **Table 1**. It should be noted that these g$A$-values were essential for reduction of intensities described in section 3.4. The *LS* percentage compositions of the observed levels from the present calculations are given in **Table 2**. They are slightly different from those given in ref. [9]. The ordering of doublet levels of the 5s5p4f configuration with ($^3$P°) and ($^1$P°) parentage was interchanged in our LSF due to inter and intra-configuration interactions, particularly, after the insertion of the $4f^25s$ configuration. As pointed out by Gayasov and Joshi, the inclusion of $4f^25s$ is important for better fitting of the levels of the 5s5p4f configuration.

**Table 3. LSF parameters ($cm^{-1}$) for Cs VII**

| Configuration[a] | Parameters[a] | LSF[a] | Unc.[b] | Index[c] | HF[a] | LSF/HF[a] |
|---|---|---|---|---|---|---|
| Odd parity | | | | | | |
| $5s^25p$ | *Eav* | 18356.20 | 249 | | 18080.40 | 1.0153 |
| | ζ(5p) | 13310.40 | 103 | 1 | 12348.80 | 1.0779 |
| $5s^26p$ | *Eav* | 331753.00 | 235 | 2 | 332208.20 | 0.9986 |
| | ζ(6p) | 4634.90 | 36 | 1 | 4300.10 | 1.0779 |
| $5s^27p$ | *Eav* | 451218.30 | 320 | 2 | 459714.30 | 0.9815 |
| | ζ(7p) | 2298.90 | 18 | 1 | 2132.80 | 1.0779 |
| $5s^28p$ | *Eav* | 516950.50 | 367 | 2 | 526684.20 | 0.9815 |
| | ζ(8p) | 1316.50 | 10 | 1 | 1221.40 | 1.0779 |
| $5s^29p$ | *Eav* | 556115.30 | 394 | 2 | 566586.40 | 0.9815 |
| | ζ(9p) | 825.90 | 6 | 1 | 766.20 | 1.0779 |
| $5s^24f$ | *Eav* | 172115.40 | 219 | | 177768.60 | 0.9682 |
| | ζ(4f) | 345.40 | fixed | | 345.40 | 1.0000 |
| $5s^25f$ | *Eav* | 385367.80 | 271 | 3 | 388697.20 | 0.9914 |
| | ζ(5f) | 99.80 | fixed | | 99.80 | 1.0000 |
| $5s^26f$ | *Eav* | 483075.90 | 339 | 3 | 487249.40 | 0.9914 |
| | ζ(6f) | 50.10 | fixed | | 50.10 | 1.0000 |
| $5s^27f$ | *Eav* | 537637.80 | 378 | 3 | 542282.70 | 0.9914 |
| | ζ(7f) | 29.30 | fixed | | 29.30 | 1.0000 |
| $5p^3$ | *Eav* | 288109.80 | 265 | | 292256.40 | 0.9858 |
| | $F^2$(5p,5p) | 48715.00 | 1009 | 4 | 61549.77 | 0.7915 |

| Configuration[a] | Parameters[a] | LSF[a] | Unc.[b] | Index[c] | HF[a] | LSF/HF[a] |
|---|---|---|---|---|---|---|
| | $\alpha$(5p) | 0.00 | fixed | | 0.00 | 0.0000 |
| | $\zeta$(5p) | 13134.00 | 102 | 1 | 12185.20 | 1.0779 |
| 5s5p5d | *Eav* | 331708.60 | 83 | | 330072.70 | 1.0050 |
| | $\zeta$(5p) | 13443.00 | 104 | 1 | 12471.80 | 1.0779 |
| | $\zeta$(5d) | 992.70 | fixed | | 992.70 | 1.0000 |
| | $F^2$(5p,5d) | 41817.70 | 729 | | 51772.24 | 0.8077 |
| | $G^1$(5s,5p) | 51605.50 | 447 | | 80524.40 | 0.6409 |
| | $G^2$(5s,5d) | 36348.20 | 781 | 5 | 41392.13 | 0.8781 |
| | $G^1$(5p,5d) | 47358.80 | 421 | 6 | 61147.07 | 0.7745 |
| | $G^3$(5p,5d) | 30217.60 | 269 | 6 | 39015.20 | 0.7745 |
| 5s5p6d | *Eav* | 527472.20 | fixed | | 527472.20 | 1.0000 |
| | $\zeta$(5p) | 14000.10 | 109 | 1 | 12988.70 | 1.0779 |
| | $G^2$(5s,6d) | 8077.20 | 174 | 5 | 9198.00 | 0.8781 |
| | $G^1$(5p,6d) | 7651.80 | 68 | 6 | 9879.60 | 0.7745 |
| | $G^3$(5p,6d) | 5857.90 | 52 | 6 | 7563.33 | 0.7745 |
| 5s5p6s | *Eav* | 405791.40 | 123 | | 401630.90 | 1.0104 |
| | $\zeta$(5p) | 13838.80 | 108 | 1 | 12839.00 | 1.0779 |
| | $G^1$(6s,5p) | 60202.40 | 469 | 7 | 81138.27 | 0.7420 |
| | $G^0$(6s,6s) | 4315.70 | 188 | 8 | 5295.60 | 0.8150 |
| | $G^1$(5p,6s) | 5858.10 | 46 | 7 | 7895.33 | 0.7420 |
| 5s5p7s | *Eav* | 555815.70 | fixed | | 555815.70 | 1.0000 |
| | $\zeta$(5p) | 14044.70 | 109 | 1 | 13030.10 | 1.0779 |
| | $G^0$(7s,6s) | 1474.90 | 64 | 8 | 1809.73 | 0.8150 |
| | $G^1$(5p,6s) | 1937.10 | 15 | 7 | 2610.80 | 0.7420 |
| 5s5d4f | *Eav* | 474788.90 | fixed | | 474788.90 | 1.0000 |
| $5p^2 4f$ | *Eav* | 428685.40 | fixed | | 428685.40 | 1.0000 |
| | $F^2$(5p,5p) | 49156.70 | 1019 | 4 | 60694.71 | 0.8099 |
| | $\zeta$(5p) | 12566.70 | 98 | 1 | 11658.80 | 1.0779 |
| $5p^2 5f$ | *Eav* | 651882.10 | fixed | | 651882.10 | 1.0000 |
| | $F^2$(5p,5p) | 50614.40 | 1049 | 4 | 62494.47 | 0.8099 |
| | $\zeta$(5p) | 13723.60 | 107 | 1 | 12732.20 | 1.0779 |
| $5p^2 6f$ | *Eav* | 752509.30 | fixed | | 752509.30 | 1.0000 |
| | $F^2$(5p,5p) | 50910.70 | 1055 | 4 | 62860.35 | 0.8099 |
| | $\zeta$(5p) | 13917.40 | 108 | 1 | 12912.00 | 1.0779 |
| $5p^2 7f$ | *Eav* | 808366.60 | fixed | | 808366.60 | 1.0000 |
| | $F^2$(5p,5p) | 51010.20 | 1057 | 4 | 62983.18 | 0.8099 |
| | $\zeta$(5p) | 13990.50 | 109 | 1 | 12979.80 | 1.0779 |
| $5p^2 6p$ | *Eav* | 596507.20 | fixed | | 596507.20 | 1.0000 |
| | $F^2$(5p,5p) | 50838.50 | 1053 | 4 | 62771.18 | 0.8099 |
| | $\zeta$(5p) | 13914.70 | 108 | 1 | 12909.50 | 1.0779 |
| | $\zeta$(6p) | 4526.50 | 35 | 1 | 4305.70 | 1.0513 |
| $5p^2 7p$ | *Eav* | 725541.80 | fixed | | 725541.80 | 1.0000 |
| | $F^2$(5p,5p) | 50986.50 | 1056 | 4 | 62953.88 | 0.8099 |
| | $\zeta$(5p) | 13996.50 | 109 | 1 | 12985.30 | 1.0779 |
| | $\zeta$(7p) | 2257.90 | 18 | 1 | 2147.80 | 1.0513 |
| $5p^2 8p$ | *Eav* | 793102.00 | fixed | | 793102.00 | 1.0000 |
| | $F^2$(5p,5p) | 51048.60 | 1058 | 4 | 63030.59 | 0.8099 |
| | $\zeta$(5p) | 14032.60 | 109 | 1 | 13018.80 | 1.0779 |
| | $\zeta$(8p) | 1289.70 | 10 | 1 | 1226.80 | 1.0513 |
| $5p^2 9p$ | *Eav* | 833262.20 | fixed | | 833262.20 | 1.0000 |
| | $F^2$(5p,5p) | 51078.20 | 1058 | 4 | 63067.18 | 0.8099 |
| | $\zeta$(5p) | 14050.50 | 109 | 1 | 13035.40 | 1.0779 |
| | $\zeta$(9p) | 808.90 | 6 | 1 | 769.40 | 1.0513 |
| Configuration interactions parameters | | | | | | |
| $5s^2 5p$-$5p^3$ | $R_d^1$(5s,5s,5p,5p) | 56509.50 | 697 | 9 | 80086.70 | 0.7056 |

| Configuration[a] | Parameters[a] | LSF[a] | Unc.[b] | Index[c] | HF[a] | LSF/HF[a] |
|---|---|---|---|---|---|---|
| $5s^25p$-5s5p5d | $R_d^1$(5s,5p,5p,5d) | 48358.90 | 596 | 9 | 68535.30 | 0.7056 |
| | $R_e^2$(5s,5p,5p,5d) | 35690.40 | 440 | 9 | 50581.30 | 0.7056 |
| $5s^25p$-5s5p6s | $R_d^0$(5s,5s,5s,6s) | 2817.90 | 35 | 9 | 3993.60 | 0.7056 |
| | $R_d^1$(5s,5p,5s,5p) | -68.20 | -1 | 9 | -96.70 | 0.7055 |
| | $R_e^0$(5s,5p,5p,5s) | 42.10 | 1 | 9 | 59.70 | 0.7048 |
| $5s^26p$-5s5p5d | $R_d^1$(5s,6p,5p,5d) | -2557.60 | -32 | 9 | -3624.70 | 0.7056 |
| | $R_e^2$(5s,6p,5p,5d) | 4719.20 | 58 | 9 | 6688.10 | 0.7056 |
| $5s^26p$-5s5p6s | $R_d^1$(5s,6p,5p,6s) | 27245.40 | 336 | 9 | 38612.80 | 0.7056 |
| | $R_e^0$(5s,6p,5p,6s) | 3564.00 | 44 | 9 | 5050.90 | 0.7056 |
| $5s^24f$-5s5p5d | $R_d^2$(4f,5s,5p,5d) | -24386.70 | -301 | 9 | -34561.50 | 0.7056 |
| | $R_e^1$(4f,5s,5p,5d) | -26223.30 | -323 | 9 | -37164.30 | 0.7056 |
| $5s^25f$-5s5p5d | $R_d^1$(5s,5f,5p,5d) | 23379.00 | 288 | 9 | 33133.20 | 0.7056 |
| | $R_e^2$(5s,5f,5p,5d) | 9749.40 | 120 | 9 | 13817.10 | 0.7056 |
| $5p^3$-5s5p5d | $R_d^1$(5p,5p,5s,5d) | 48155.20 | 594 | 9 | 68246.70 | 0.7056 |
| $5p^3$-5s5p6s | $R_d^1$(5p,5p,5s,6s) | -11.50 | 0 | 9 | -16.30 | 0.7070 |
| 5s5p5d-5s5p6s | $R_d^2$(5p,5d,5p,5s) | -9025.80 | -111 | 9 | -12791.60 | 0.7056 |
| | $R_e^1$(5p,5d,5p,5s) | -3149.10 | -39 | 9 | -4462.90 | 0.7056 |
| Standard deviation (SD) = 303 $cm^{-1}$ | | | | | | |
| even parity | | | | | | |
| $5s^26s$ | *Eav* | 279901.70 | 291 | | 280154.50 | 0.9991 |
| $5s^27s$ | *Eav* | 434262.90 | fixed | | 434262.90 | 1.0000 |
| $5s^28s$ | *Eav* | 512265.20 | fixed | | 512265.20 | 1.0000 |
| $5s^29s$ | *Eav* | 557603.90 | fixed | | 557603.90 | 1.0000 |
| $5s^25d$ | *Eav* | 210015.60 | 258 | | 212599.40 | 0.9878 |
| | $\zeta$(5d) | 1283.60 | 179 | 1 | 973.60 | 1.3184 |
| $5s^26d$ | *Eav* | 406872.40 | fixed | | 406872.40 | 1.0000 |
| | $\zeta$(6d) | 550.50 | 77 | 1 | 417.50 | 1.3186 |
| $5s^27d$ | *Eav* | 498083.70 | fixed | | 498083.70 | 1.0000 |
| | $\zeta$(7d) | 293.50 | 41 | 1 | 222.60 | 1.3185 |
| $5s^28d$ | *Eav* | 549187.00 | fixed | | 549187.00 | 1.0000 |
| | $\zeta$(8d) | 176.40 | 25 | 1 | 133.80 | 1.3184 |
| $5s^29d$ | *Eav* | 580950.20 | fixed | | 580950.20 | 1.0000 |
| | $\zeta$(9d) | 114.70 | 16 | 1 | 87.00 | 1.3184 |
| $5s^25g$ | *Eav* | 456153.40 | fixed | | 456153.40 | 1.0000 |
| | $\zeta$(5g) | 2.40 | fixed | | 2.40 | 1.0000 |
| $5s^26g$ | *Eav* | 523875.00 | fixed | | 523875.00 | 1.0000 |
| | $\zeta$(6g) | 1.50 | fixed | | 1.50 | 1.0000 |
| $5s^27g$ | *Eav* | 564456.80 | fixed | | 564456.80 | 1.0000 |
| | $\zeta$(7g) | 0.94 | fixed | | 0.94 | 1.0000 |
| $5s5p^2$ | *Eav* | 146889.70 | 133 | | 143667.90 | 1.0224 |
| | $F^2$(5p,5p) | 51993.70 | 900 | 2 | 61560.35 | 0.8446 |
| | $\alpha$(5p) | 0.00 | fixed | | 0.00 | 0.0000 |
| | $\zeta$(5p) | 13363.90 | 154 | 3 | 12263.40 | 1.0897 |
| | $G^1$(5s,5p) | 59525.30 | 348 | | 80087.73 | 0.7433 |
| $5s5d^2$ | *Eav* | 524396.30 | fixed | | 524396.30 | 1.0000 |
| | $\zeta$(5d) | 1331.10 | 186 | 1 | 1009.60 | 1.3184 |
| 5s5p4f | *Eav* | 290781.00 | 75 | | 291898.50 | 0.9962 |
| | $\zeta$(4f) | 352.00 | fixed | | 352.00 | 1.0000 |
| | $\zeta$(5p) | 12803.60 | 148 | 3 | 11749.20 | 1.0897 |
| | $F^2$(4f,5p) | 43120.10 | 766 | | 52178.94 | 0.8264 |
| | $G^3$(4f,5s) | 32055.90 | 1250 | | 35026.93 | 0.9152 |
| | $G^2$(4f,5p) | 32095.50 | 1315 | | 33280.67 | 0.9644 |
| | $G^4$(4f,5p) | 17079.00 | 1525 | | 25087.07 | 0.6808 |
| | $G^1$(5s,5p) | 52195.70 | 268 | 4 | 78949.87 | 0.6611 |
| 5s5p5f | *Eav* | 508633.40 | fixed | | 508633.40 | 1.0000 |

| Configuration[a] | Parameters[a] | LSF[a] | Unc.[b] | Index[c] | HF[a] | LSF/HF[a] |
|---|---|---|---|---|---|---|
| | ζ(5p) | 13968.90 | 161 | 3 | 12818.60 | 1.0897 |
| | $G^1$(5s,5p) | 53693.70 | 276 | 4 | 81215.73 | 0.6611 |
| 5s6p4f | *Eav* | 590990.90 | fixed | | 590990.90 | 1.0000 |
| | ζ(6p) | 4654.80 | 54 | 3 | 4271.50 | 1.0897 |
| | $G^1$(5s,6p) | 5744.60 | 30 | 4 | 8689.20 | 0.6611 |
| 5s5p6p | *Eav* | 452657.00 | fixed | | 452657.00 | 1.0000 |
| | ζ(5p) | 14166.90 | 163 | 3 | 13000.30 | 1.0897 |
| | ζ(6p) | 4681.60 | 54 | 3 | 4296.10 | 1.0897 |
| | $G^1$(5s,5p) | 53937.30 | 277 | 4 | 81584.27 | 0.6611 |
| | $G^1$(5s,6p) | 5529.00 | 28 | 4 | 8363.07 | 0.6611 |
| 5s5p7p | *Eav* | 580934.80 | fixed | | 580934.80 | 1.0000 |
| | ζ(5p) | 14248.00 | 164 | 3 | 13074.70 | 1.0897 |
| | ζ(7p) | 2331.20 | 27 | 3 | 2139.20 | 1.0898 |
| | $G^1$(5s,5p) | 54061.30 | 278 | 4 | 81771.87 | 0.6611 |
| | $G^1$(5s,7p) | 2064.30 | 11 | 4 | 3122.40 | 0.6611 |
| $5p^26s$ | *Eav* | 546667.30 | fixed | | 546667.30 | 1.0000 |
| | $F^2$(5p,5p) | 52726.30 | 912 | 2 | 62427.65 | 0.8446 |
| | ζ(5p) | 13899.70 | 160 | 3 | 12755.10 | 1.0897 |
| | $G^1$(5p,6s) | 5270.60 | 27 | 4 | 7972.13 | 0.6611 |
| $5p^27s$ | *Eav* | 700796.50 | fixed | | 700796.50 | 1.0000 |
| | $F^2$(5p,5p) | 53121.30 | 919 | 2 | 62895.41 | 0.8446 |
| | ζ(5p) | 14103.00 | 163 | 3 | 12941.60 | 1.0897 |
| | $G^1$(5p,7s) | 1732.40 | 9 | 4 | 2620.40 | 0.6611 |
| $5p^28s$ | *Eav* | 779035.50 | fixed | | 779035.50 | 1.0000 |
| | $F^2$(5p,5p) | 53218.10 | 921 | 2 | 63010.00 | 0.8446 |
| | ζ(5p) | 14165.90 | 163 | 3 | 12999.40 | 1.0897 |
| | $G^1$(5p,8s) | 828.90 | 4 | 4 | 1253.73 | 0.6611 |
| $5p^29s$ | *Eav* | 824486.30 | fixed | | 824486.30 | 1.0000 |
| | $F^2$(5p,5p) | 53258.10 | 922 | 2 | 63057.29 | 0.8446 |
| | ζ(5p) | 14193.70 | 164 | 3 | 13024.90 | 1.0897 |
| | $G^1$(5p,9s) | 471.10 | 2 | 4 | 712.53 | 0.6612 |
| $5p^25d$ | *Eav* | 470629.80 | fixed | | 470629.80 | 1.0000 |
| | $F^2$(5p,5p) | 52290.80 | 905 | 2 | 61912.12 | 0.8446 |
| | ζ(5p) | 13497.30 | 156 | 3 | 12385.80 | 1.0897 |
| | ζ(5d) | 1335.10 | 186 | 1 | 1012.60 | 1.3185 |
| $5p^26d$ | *Eav* | 671480.70 | fixed | | 671480.70 | 1.0000 |
| | $F^2$(5p,5p) | 53086.70 | 919 | 2 | 62854.35 | 0.8446 |
| | ζ(5p) | 14058.40 | 162 | 3 | 12900.70 | 1.0897 |
| | ζ(6d) | 554.10 | 77 | 1 | 420.30 | 1.3183 |
| $5p^27d$ | *Eav* | 764069.10 | fixed | | 764069.10 | 1.0000 |
| | $F^2$(5p,5p) | 53195.70 | 921 | 2 | 62983.41 | 0.8446 |
| | ζ(5p) | 14148.10 | 163 | 3 | 12983.00 | 1.0897 |
| | ζ(7d) | 294.80 | 41 | 1 | 223.60 | 1.3184 |
| $5p^28d$ | *Eav* | 815653.20 | fixed | | 815653.20 | 1.0000 |
| | $F^2$(5p,5p) | 53244.80 | 921 | 2 | 63041.65 | 0.8446 |
| | ζ(5p) | 14184.50 | 163 | 3 | 13016.40 | 1.0897 |
| | ζ(8d) | 177.10 | 25 | 1 | 134.30 | 1.3187 |
| $5p^29d$ | *Eav* | 847643.00 | fixed | | 847643.00 | 1.0000 |
| | $F^2$(5p,5p) | 53270.80 | 922 | 2 | 63072.35 | 0.8446 |
| | ζ(5p) | 14203.00 | 164 | 3 | 13033.40 | 1.0897 |
| | ζ(9d) | 115.10 | 16 | 1 | 87.30 | 1.3184 |
| $5p^25g$ | *Eav* | 721771.70 | fixed | | 721771.70 | 1.0000 |
| | $F^2$(5p,5p) | 53106.80 | 919 | 2 | 62878.24 | 0.8446 |
| | ζ(5p) | 14123.00 | 163 | 3 | 12960.00 | 1.0897 |
| $5p^26g$ | *Eav* | 790137.80 | fixed | | 790137.80 | 1.0000 |

| Configuration[a] | Parameters[a] | LSF[a] | Unc.[b] | Index[c] | HF[a] | LSF/HF[a] |
|---|---|---|---|---|---|---|
| | $F^2$(5p,5p) | 53191.20 | 921 | 2 | 62978.12 | 0.8446 |
| | $\zeta$(5p) | 14160.20 | 163 | 3 | 12994.10 | 1.0897 |
| $5p^27g$ | *Eav* | 831009.50 | fixed | | 831009.50 | 1.0000 |
| | $F^2$(5p,5p) | 53240.00 | 921 | 2 | 63035.88 | 0.8446 |
| | $\zeta$(5p) | 14186.60 | 163 | 3 | 13018.40 | 1.0897 |
| $4f^25s$ | *Eav* | 453041.70 | fixed | | 453041.70 | 1.0000 |
| Configuration interactions parameters | | | | | | |
| $5s^26s$-$5s5p^2$ | $R_d^1$(5s,6s,5p,5p) | -665.40 | -16 | 5 | -865.70 | 0.7686 |
| $5s^25d$-$5s5p^2$ | $R_d^1$(5s,5d,5p,5p) | 52178.40 | 1248 | 5 | 67892.00 | 0.7686 |
| $5s^25d$-5s5p4f | $R_d^3$(5s,5d,4f,5p) | -23261.10 | -557 | 5 | -30266.30 | 0.7685 |
| | $R_e^1$(5s,5d,4f,5p) | -27549.80 | -659 | 5 | -35846.40 | 0.7686 |
| $5s5p^2$-5s5p4f | $R_d^2$(5p,5p,4f,5p) | -32011.20 | -766 | 5 | -41651.50 | 0.7685 |
| SD = 284 $cm^{-1}$ | | | | | | |

[a] Configurations involved in the calculations and their Slater parameters with the corresponding Hartree–Fock (HF) and/or least-squares-fitted (LSF) values and their ratios.

[b] Uncertainty of each parameter represents its standard deviation (SD).

[c] Parameters in each numbered group were linked together with their ratio fixed at the HF level, other remaining configuration-interaction parameters $R^k$ even and odd configurations were fixed at 75% of the Hartree–Fock value.

For the $5s^25p$–{$5s5p^2$, $5s^2$(5d+6s)} transitions, the *gA*-values calculated with the present HFR model were compared with those we calculated with the fully-relativistic multiconfiguration Dirac-Hartree-Fock (MCDHF) method for $Cs^{6+}$ ion (see **Table 4**). In the present MCDHF calculations, only valance electron correlations were included via single-double excitations from $5s^2n\ell$ (*n* up to 8 for $\ell$=s, p, d), $5s^2$(4f+5f+5g), and $5s5p^2$ (*n* up to 8 for $\ell$=s, p, d) type of configurations. Only the main results of the present MCDHF calculation are summarized in this work, and full details of it can be found elsewhere [14]. The average fractional uncertainty indicator (*dT*) for the MCDHF calculated *gA*-values, computed using differences of the quantities in the length and velocity forms, was 15 %. A similar value was also obtained from the estimation of the corresponding line strengths (*S*-values), excluding a weak line with $S \approx 1.86\times10^{-3}$ atomic units (a.u.) at 821.094 Å. The reliability of *gA*-values in each model is indicated by their absolute cancellation factor (|CF|) for HFR and fractional uncertainty indicator (*dT*) for the MCDHF. Unlike the d*T* indicator for MCDHF model, the *gA*-values from HFR are unreliable for transitions with small cancellation factor, |CF|<0.10. The general agreement between the HFR and MCDHF results is good, i.e., within 50 % on average, but one can see a clear dependency on the magnitude of the line strength (*S*-value) that strongly correlates with discrepancies between different calculations and thus, with their uncertainties. The HFR line strengths ($S_{\mathrm{HFR}}$) with $S_{\mathrm{HFR}} \geq 0.80$ a.u. deviate from the MCDHF line strengths ($S_{\mathrm{MCDHF}}$) by 19 %, while the remaining weaker lines show a 78 % mean deviation. These mean deviations were further combined in quadrature with the 15 % uncertainty of the MCDHF model (see **Table 4**). However, one should be cautious while considering the mere internal consistency of the length and velocity forms of any theoretical model. Using it as an ultimate uncertainty indicator for the transition parameter may not always be correct. The method of evaluation of transition parameters should be more comprehensive as described in ref. [15]. Our HFR calculations are much more extensive in comparison to the MCDHF ones, and hence we consider them to be more accurate. Due to the above reasons, we do not recommend any final uncertainties for the values given in **Table 4**.

**Table 4**. Transition probabilities of selected lines from the HFR and MCDHF methods.

| $\lambda_{obs}$[a], (Å) | Intensity[a] | $gA_{HFR}$[b], ($s^{-1}$) | \|CF\|[b] | $gA_{mcdhf}$[b], ($s^{-1}$) | $dT$[b] | Unc. (%)[c] | Comment[c] |
|---|---|---|---|---|---|---|---|
| 365.834(5) | 2200 | 1.80e+10 | 0.91 | 1.28e+10 | 0.06 | 80 | |
| 393.745(5) | 6000 | 2.88e+10 | 0.91 | 2.93e+10 | 0.08 | 25 | |
| 485.449(5) | 53000 | 8.82e+10 | 0.64 | 7.92e+10 | 0.11 | 25 | |
| 527.958(5) | 36000 | 1.33e+11 | 0.76 | 1.27e+11 | 0.13 | 25 | |
| 535.860(5) | 6300 | 2.15e+10 | 0.73 | 2.06e+10 | 0.14 | 25 | |
| 557.599(5) | 14000 | 1.62e+10 | 0.30 | 1.75e+10 | 0.17 | 25 | |
| 561.197(5) | 1900 | 1.77e+09 | 0.04 | 1.45e+10 | 0.19 | | UR |
| 625.146(10) | 4700bl | 5.95e+10 | 0.55 | 5.45e+10 | 0.18 | 25 | |
| 629.679(5) | 9300 | 2.01e+10 | 0.59 | 1.86e+10 | 0.16 | 25 | |
| 632.544(5) | 7900 | 2.70e+10 | 0.65 | 1.67e+10 | 0.15 | 25 | |
| 708.370(5) | 4600 | 9.06e+09 | 0.21 | 7.67e+09 | 0.08 | 25 | |
| 720.916(5) | 900 | 9.11e+08 | 0.04 | 7.03e+08 | 0.05 | 80 | |
| 782.451(5) | 6700 | 6.08e+09 | 0.15 | 6.10e+09 | 0.05 | 25 | |
| 821.094(5) | 190 | 1.49e+08 | 0.01 | 6.82e+06 | 0.69 | | UR |
| 875.240(5) | 18 | 8.57e+06 | 0.02 | 8.73e+06 | 0.12 | 80 | |
| 959.452(5) | 590 | 2.84e+08 | 0.59 | 1.42e+08 | 0.15 | 80 | |
| 971.984(5) | 1200 | 8.13e+08 | 0.23 | 3.81e+08 | 0.05 | 80 | |
| 1054.015(5) | 130 | 8.65e+07 | 0.42 | 3.31e+07 | 0.23 | 80 | |
| 1178.596(5) | 25 | 2.90e+07 | 0.09 | 3.23e+07 | 0.02 | 80 | |

[a] Observed wavelength and line intensity from Table 1.

[b] Transition probabilities ($gA$-values) from the present HFR and MCDHF calculations (see text) rescaled experimental energies. The reliability indicator for $gA$-values from each model is represented by the absolute cancellation factor (|CF|) for HFR and fractional uncertainty indicator ($dT$) for MCDHF (see text).

[c] Relative uncertainty in percentages for each $gA$-values in column 2 and notes regarding the most unreliable (UR) $gA$-values from the HFR and/or MCDHF method, which were omitted in the comparison.

In diagnostic of astrophysical plasmas, the most important transition of Cs VII is between $^2P^o$ levels of the $5s^25p$ configuration. From our level optimization, the $^2P^o_{1/2}$–$^2P^o_{3/2}$ separation is found to be 19379.3(4) $cm^{-1}$, and the corresponding forbidden line (M1+E2) is at 5160.15(11) Å in vacuum. The data on the transition rates (M1 and E2) of this line are summarized in **Table 5**. It is worth to note that the agreement between the values from the all-order ($Z^{SDpT}$) multipole matrix elements [4] and MCDHF methods [14, 16] is excellent, within 0.04 % for the M1 transition, while the HFR value deviates by 1 % from the average of $Z^{SDpT}$ and MCDHF. All five values agree within 10 % for the E2 component of the line. A similar assessment can also be given for the lifetime data calculated by various methods. Since the M1 and E2 contributions are 0.99 % and 0.01%, respectively, the lifetime of the $5s^25p$ $^2P^o_{3/2}$ state is predominantly defined by the M1 decay rate. Thus, the agreement of the $Z^{SDpT}$ and MCDHF results for the lifetime is within 0.10 %. From these comparisons, the best values proposed for the M1 and E2 decay rates of $5s^25p\ ^2P^o_{3/2}$ state are 65.123(25) $s^{-1}$ and 0.68(5) $s^{-1}$, respectively, and for the lifetime it is 15.197(14) milliseconds.

**Table 5**. Transition rates (*A*-values in $s^{-1}$) for the astrophysically important $5s^25p\ ^2P^o$ (1/2–3/2) line.

| Theoretical Method[a] | Original work | | | Scaled values[b] | | Lifetime[c] (ms) |
|---|---|---|---|---|---|---|
| | λ, (Å) | M1 | E2 | M1 | E2 | |
| $Z^{SDpT}$, [4] | 5160 | 65.10 | 0.6371 | 65.10 | 0.6371 | 15.200 |
| MCDHF, [16] | 5161.44 | 65.10 | 0.7171 | 65.15 | 0.7180 | 15.182 |
| MCDHF, [14] | 5242.46 | 62.10 | 0.5593 | 65.12 | 0.6054 | 15.215 |
| HFR+LSF, [17] | 5160 | 65.75 | 0.7105 | 65.75 | 0.7105 | 15.047 |
| HFR+LSF, TW | 5480.95 | 54.75 | 0.5270 | 65.73 | 0.7151 | 15.051 |

[a] Theoretical method and reference for the *A*-values (M1 and E2) given in columns 3 & 4, wherein 'TW' refers to the present work.
[b] The originally reported values were scaled with respect to the experimental wavelength 5160.15 Å (in vacuum), except for those already scaled in refs. [4] and [17].
[c] Calculated lifetime of the $5s^25p\ ^2P^o_{3/2}$ excited level in milliseconds.

### 3.2. Analysis of the spectrum

Our analysis on Cs VII has begun with verification of lines and levels reported by Gayasov and Joshi [9]. We confirmed all of them except a few weak lines. Nonetheless, the work of Gayasov and Joshi is described in detail in the section below as it is one of the prime sources for the present work.

#### 3.2.1. Measurements of Gayasov and Joshi

The spectrograms used by Gayasov and Joshi [9] for the analysis of Cs VII were recorded on the same Antigonish 3-m normal incidence vacuum spectrograph in the wavelength region 300–1240 Å using the triggered spark sources briefly described in Section 2 and also in ref. [12]. The exposures were taken either on Kodak SWR or 101–05 plates. The relative positions of spectral lines have been measured on an automatic Grant comparator, which provides better position accuracy for lines than the Abbe comparator. The wavelength reduction was carried out with the help of internal standards provided by impurities of C, N, O, and Al lines. The line intensities reported by Gayasov and Joshi on a scale from 2 to 1000 were obtained from an optical densitometer (see Section 3.4).

Gayasov and Joshi could easily segregate the lines of Cs VII in their spectrograms as they had already studied several spectra of cesium ions (Cs IV to Cs IX) in their laboratory (see Ref. [9] and references therein). This helped them to revise the work of Kaufman and Sugar [7] and extend it further. The lines of $5s^25p\ ^2P^o$–$5s5p^2\ ^4P$ transitions were first observed by Tauheed et al. [8]. Later, their improved wavelengths were given by Gayasov and Joshi. A total of 184 lines were observed by Gayasov and Joshi. They belong to the $5s^25p$–{$5s5p^2$, $5s^2(5d+6s)$, 5s5p4f}, $5s5p^2$–{$5s^26p$,5s5p(5d+6s), $5p^3$}, $5s^25d$–{$5s^2(5f+6p)$, 5s5p(5d+6s)}, $5s^26s$–5s5p6s, and $5s^24f$–

5s5p4f transition arrays. All levels of the configurations involved were found except two out of twenty-three levels of 5s5p5d and 5s5p4f (7 out of 24 levels). All but a few weak lines reported by Gayasov and Joshi appeared on our plates with line characteristics of Cs VII. Therefore, we accepted all the lines reported by Gayasov and Joshi and retained them in the final level optimization (see section 3.3). An additional test was also performed by us using Boltzmann plots. In this test, the observed line intensities were found to be strongly correlated with transition rates (see section 3.4). It should be noted that our extended HFR calculations and the subsequently made LSF (see section 3.1) are superior to those made by Gayasov and Joshi, as we considered an extensive sets of almost all interacting configurations in the calculations. Additionally, we linked all similar parameters in the LSF calculations resulting in an improved value for the parameters involved (see **Table 3**).

### 3.3. Optimization of energy levels

After confirming almost all the lines identified by Gayasov and Joshi [9], we performed the optimization of energy levels using the LOPT code [18]. In the initial stage of the optimization, only wavelengths of Cs VII lines reported by Gayasov and Joshi with an uncertainty of 0.005 Å were included in the input, and those from our plates were ignored due to their inferior accuracy. Using these optimized energy levels and their Ritz wavelengths; we found 12 previously unobserved lines (with strong TPs) on our spectrograms and added them to the line list, as well as one line masked by another species. The final line list of Cs VII is given in **Table 1**, while the optimized energy levels with their uncertainties are given in **Table 2**. Our listed energy values can immediately be compared with those reported by Gayasov and Joshi [9] or with those reproduced from ref. [9] by Sansonetti [10] for the NIST ASD [11] with their uncertainties. There are no significant differences between them, but the uncertainties of energy levels given in ref. [10] were overestimated. In other words, we retained their required numerical precision and uncertainties from the same input data, ref. [9]. We used the present optimized energy levels to compute accurate wavelengths of several possibly observable lines. We also provide the transition probabilities for these lines, which could be important for detailed plasma models.

### 3.4. Reduction of line intensities

A visual estimate of blackening of the photographic emulsion is one of the commonly given quantities to represent the relative line intensities. With rare exceptions optical densitometers are commonly used for this purpose. Thus, most often the observers give only rough estimates of relative line intensities. Nevertheless, the line intensities are vital for proper identification of lines. Generally, different observers use different scales to express the relative line intensities. Moreover, efficiency of spectrographs (including their optics), photographic emulsions, and excitation conditions of light sources are also different. Bringing such heterogeneous intensities to a common scale is a non-trivial problem. If several sets of intensities observed at different conditions are available, each observation should be analysed separately through a method using the Boltzmann equation to approximate the levels populations together with theoretically obtained transition probabilities (*gA*-values). This method was successfully used in analyses of many similar spectra [19–22]. Once an approximate excitation temperature is determined for the light source used then spectral response functions (in terms of $\lambda$) of the instrument (including the detector) are easily determined by comparing the observed and modelled intensities, and spectral variations of the efficiency of the instrument are removed from the observed intensities. Subsequently, intensities observed with another instrument or those from the same instrument under different excitation condition can be reduced to a global uniform scale with a common effective excitation temperature ($T_{\text{eff}}$) chosen from one of the observations (see further details in ref. [19–22]).

As briefly described in section 3.2.1, Gayasov and Joshi [9] measured the relative line intensities of Cs VII lines with an optical densitometer. The photographic emulsion characteristic curve was used by them without considering any $\lambda$-dependency of the emulsion. However, the background in the spectrogram was approximated by a spline curve and was removed. Their final intensities are given on a linear scale from 2 to 1000. We use these intensities together with *gA*-values obtained from our HFR calculations described above (see section 3.1) to reduce them to a uniform scale corresponding to $T_{\text{eff}}$ = 6.6 eV, as derived from the Boltzmann plot shown in **Figure 1**. In this figure, the corrected intensities given in **Table 1** were used. The logarithmic intensity correction functions varying with wavelength ($\lambda$-dependency) for the intensities reported in Ref. [9] are given in **Figure 2**. A few data points deviating too much from the fitted curves were removed from these plots. The corresponding lines are marked with a special note in **Table 1**. These large deviations can be explained by unidentified blending in the spectra observed in Ref. [9].

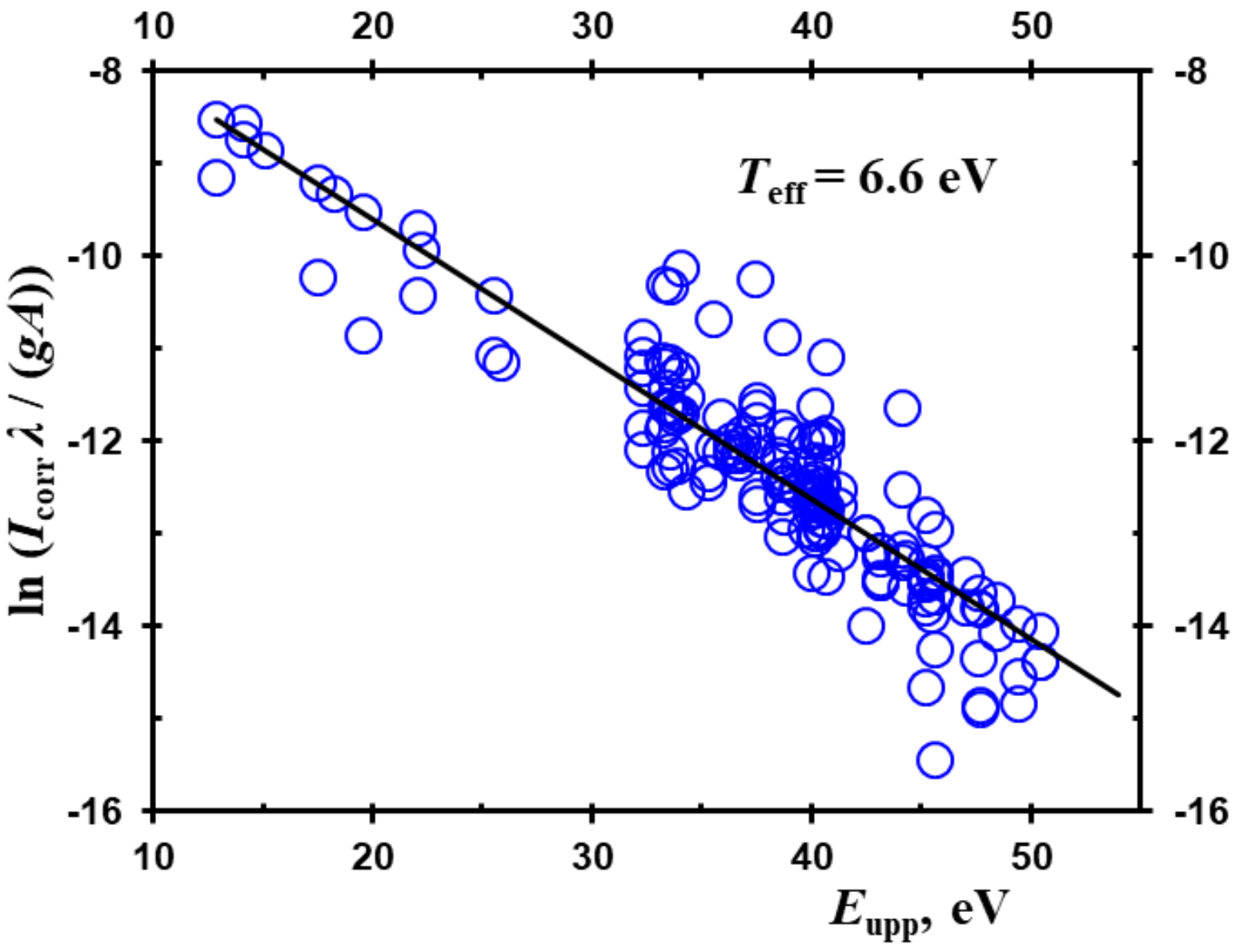

Figure 1. Boltzmann plot for the observed line intensities of Gayasov and Joshi [9]. The effective excitation temperature, $T_{eff}$ = 6.6 eV, was derived from the corrected intensities ($I_{corr}$), which were obtained after the removal of the wavelength-dependent correction from the originally observed line intensities (see figure 2).

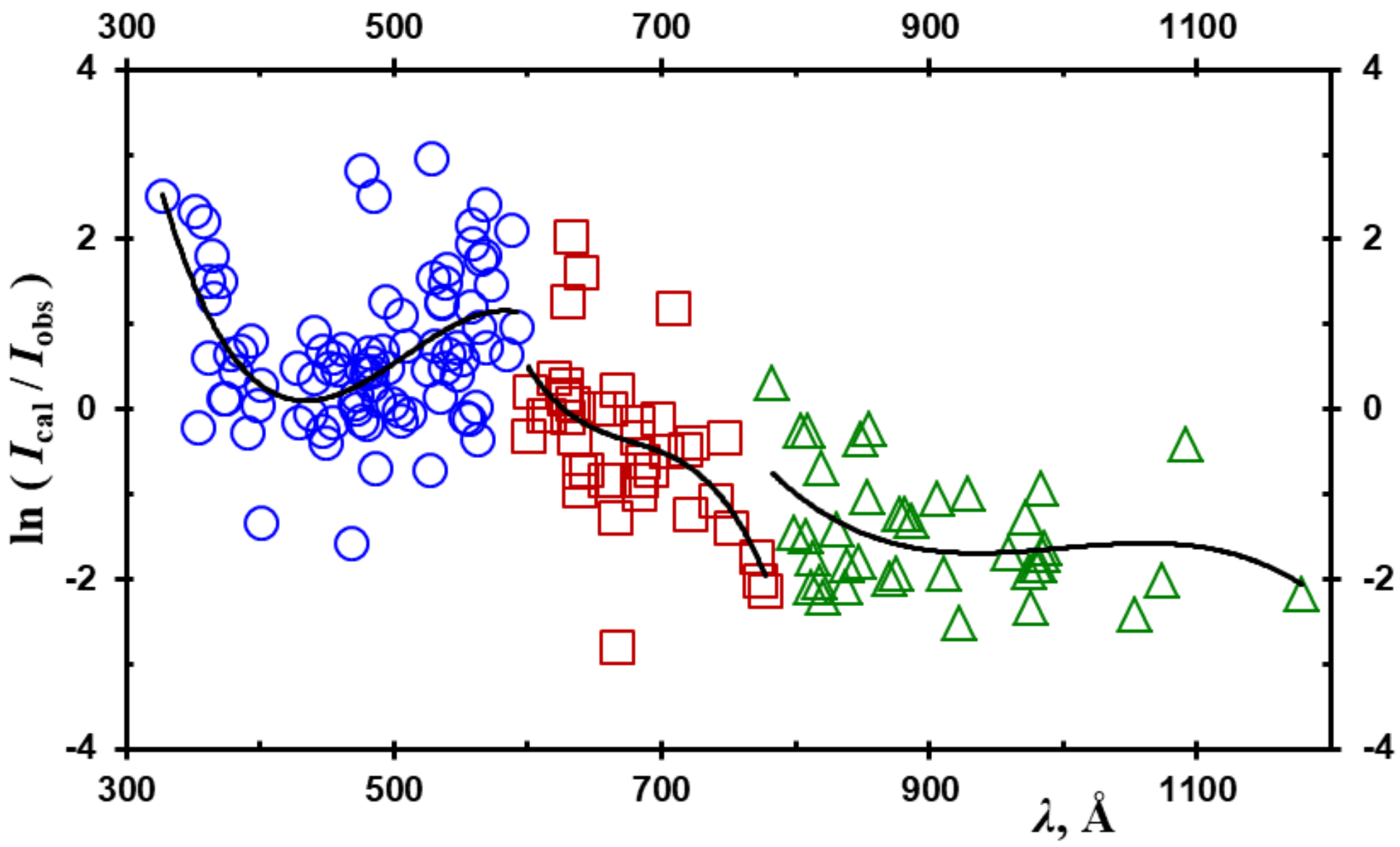

Figure 2. Logarithmic intensity correction functions for the observation ($I_{obs}$) of Gayasov and Joshi [9], which smoothly vary with the wavelength ($\lambda$ in Å) in three different regions. The calculated line intensities are obtained with formula $I_{cal} = (gA/\lambda)\exp(-E_{upp}/T_{eff} + C_0)$. The solid lines represent cubic polynomial functions fitted to the data.

Similar procedures were followed for the intensities of Cs VII observed on our spectrograms, which were taken from two plates recorded at different experimental conditions on the same instrument. The first plate covers the wavelength region 300–890 Å, while second one is in the 730–1240 Å range of wavelength. The final $T_{eff}$ values derived for them were 10.4 eV and 6.8 eV, respectively. The corrected intensities were then rescaled to the global uniform $T_{eff}$ = 6.6 eV selected from the spectrum of Gayasov and Joshi. It should be noted that our originally recorded intensities are visual estimates of photographic blackening, which inherently tend to have large variations in comparison with those from an optical densitometer. For this reason, no intensity averaging has been made in the final line intensities given in **Table 1**. Nonetheless, this modelling was helpful in determining the line intensities of 12 newly identified lines on our spectrograms.

## 4. Conclusions

The earlier reported work on the spectrum of Cs VII by Gayasov and Joshi [9] has been re-investigated thoroughly using supplementary spectrograms of cesium recorded on a 3 m normal incidence vacuum spectrograph in the 300–1240 Å wavelength region. We confirmed the lines and levels reported earlier by Gayasov and Joshi and found twelve new Cs VII lines on our spectrograms. Four of these new lines establish two previously unknown levels of 5s5p4f and a third level of the 5s5p5d configurations. In this work, besides providing newly optimized energy levels of Cs VII with their uncertainties, as well as observed and Ritz wavelengths with uncertainties, we also give transition probabilities and uniformly scaled intensities of Cs VII lines. For a few selected lines, the transition probabilities from our HFR calculations were evaluated by comparison with those from the MCDHF method. A total of 196 lines attributed to 197 transitions enabled us to optimize the energy values of 72 levels in the Cs VII spectrum. The line list of Cs VII was further enhanced with 140 possibly observable lines, for which Ritz wavelengths are provided along with their *gA*-values.

**Acknowledgments**

AH is thankful to the University Grants Commission (UGC), India for providing the financial support to carry out this work. A. Tauheed would like to thank St. Francis Xavier University, Antigonish (Canada) and late Prof. Y. N. Joshi for providing local hospitality during the recording of cesium spectra. We are also grateful to Dr. Alexander Kramida of NIST for providing the modified version of Cowan’s code 2019 (Private communication).

**References**

[1] Indelicato P. QED tests with highly charged ions. J Phys B:At Mol Opt Phys 2019;52:232001. https://doi.org/10.1088/1361-6455/ab42c9.

[2] Kozlov MG, Safronova MS, et al. Highly charged ions: Optical clocks and applications in fundamental physics. Rev Mod Phys 2018;90:045005. https://doi.org/10.1103/RevModPhys.90.045005.

[3] Safronova MS, Dzuba VA, et al. Highly Charged Ions for Atomic Clocks, Quantum Information, and Search for α variation. Phys Rev Lett 2014;113:030801. https://doi.org/10.1103/PhysRevLett.113.030801.

[4] Safronova MS, Dzuba VA, et al. Highly charged Ag-like and In-like ions for the development of atomic clocks and the search for α variation. Phys Rev A 2014;90:042513. https://doi.org/10.1103/PhysRevA.90.042513.

[5] Hinkley N, Sherman JA, Phillips NB, et al. An Atomic Clock with $10^{-18}$ Instability. Science 2013;341:1215. https://doi.org/10.1126/science.1240420.

[6] Bloom BJ, Nicholson TL, Williams JR, et al. An optical lattice clock with accuracy and stability at the $10^{-18}$ level. Nature 2014;71:506. https://doi.org/10.1038/nature12941.

[7] Kaufman V, Sugar J. In I isoelectronic sequence: wavelengths and energy levels for Xe VI through La IX. J Opt Soc 1987;4:1924. https://doi.org/10.1364/JOSAB.4.001924

[8] Tauheed A, Joshi YN, Pinnington EH. The $5s^25p\ ^2P$-$5s5p^2\ ^4P$ intercombination lines in the In I isoelectronic sequence from Sb III to La IX. J Phys B: At Mol Opt Phys 1992;25:561. https://doi.org/10.1088/0953-4075/25/22/002.

[9] Gayasov R, Joshi YN. Revised and Extended Analysis of Six-times Ionized Cesium: Cs VII. J Phys Scr 1999;60:312. https://doi.org/10.1238/physica.regular.060a00312.

[10] Sansonetti JE. Wavelengths, Transition Probabilities, and Energy Levels for the Spectra of Cesium (CsI–Cs LV). J Phys Chem Ref Data 2009;38:720. https://doi.org/10.1063/1.3132702.

[11] Kramida A, Ralchenko Y, Reader J, and NIST ASD Team; 2020. NIST Atomic Spectra Database (ver. 5.7.1), [online]. Available: https://physics.nist.gov/asd.
[12] Gayasov R, Joshi YN. Revised and extended analysis of seven-timesionized cesium: CsVIII. J Opt Soc 1999;16:1280. https://doi.org/10.1364/JOSAB.16.001280.
[13] Cowan RD. The Theory of Atomic Structure and Spectra, Berkeley, CA, USA: University of California Press;1981. Cowan Code package for Windows by Kramida A. A suite of atomic structure codes originally developed by R. D. Cowan adapted for windows-based personal computers. NIST Public DATA Repository;2019. https://doi.org/10.18434/T4/1502500.
[14] Wajid A, Husain A, Jabeen S, Tauheed A. The Multi-configuration Dirac-Hartree-Fock calculations for Cs VII. Submitted to J At Mol Cond Nano Phys.
[15] Hibbert, A. Successes and Difficulties in Calculating Atomic Oscillator Strengths and Transition Rates. Galaxies 2018;6;77. https://doi.org/10.3390/galaxies6030077
[16] Wang HW, Zhang L, Jiang G, et al. The energy levels and transition properties of In-like ions. Ind J Phys 2018;92:137. https://doi.org/10.1007/s12648-017-1094-z.
[17] Biemont E, Hansen JE, et al. Forbidden transitions of astrophysical interest in the $5p^k$ ($k$=1–5) configurations. Astron Astrophys Suppl Ser 1995;111:333. http://adsabs.harvard.edu/abs/1995A&AS..111..333B.
[18] Kramida A. The program LOPT for least-squares optimization of energy levels. Comput Phys Commun 2011;182:419. https://doi.org/10.1016/j.cpc.2010.09.019.
[19] Kramida A. Critical Evaluation of Data on Atomic Energy Levels, Wavelengths, and Transition Probabilities. Fusion Sci Technol 2013;63:313. https://doi.org/10.13182/FST13-A16437.
[20] Kramida A. Critically Evaluated Energy Levels and Spectral Lines of Singly Ionized Indium (In II). J Res Natl Inst Tech 2013;118:52. https://doi.org/10.6028/jres.118.004.
[21] Kramida A. A Critical Compilation of Energy Levels, Spectral Lines, and Transition Probabilities of Singly Ionized Silver, Ag II. J Res Natl Inst Tech 2013;118:168. https://doi.org/10.6028/jres.118.009.
[22] Haris K, Kramida A, Tauheed A. Extended and revised analysis of singly ionized tin: Sn II. Phy Scr 2014;89:115403. https://doi.org/10.1088/0031-8949/89/11/115403.